\documentclass[aps,pre,twocolumn,superscriptaddress
]{revtex4-2}

\usepackage{graphicx}
\usepackage{dcolumn}
\usepackage{bm}

\usepackage{xcolor}
\usepackage{comment}
\usepackage{arydshln}
\usepackage{afterpage}
\usepackage{amsmath}
\usepackage{amssymb}

\begin{document}

\preprint{APS/123-QED}

\title{Mechanism-based metamaterials with microstructurally invariant shape-change}




\author{Yingchao Peng}
\affiliation{Aerospace and Mechanical Engineering, University of Southern California, Los Angeles, CA 90014, USA}
\author{Asifur Rahman}%
\affiliation{Civil Engineering, Stony Brook University, Stony Brook, NY 11794, USA}
\author{Paolo Celli}%
\affiliation{Civil Engineering, Stony Brook University, Stony Brook, NY 11794, USA}
\author{Paul Plucinsky}
\email{plucinsk@usc.edu}
\affiliation{Aerospace and Mechanical Engineering, University of Southern California, Los Angeles, CA 90014, USA}%
\date{\today}
 




\date{\today}

\begin{abstract}


Metamaterials with  floppy modes called mechanisms 
are a burgeoning  template for shape-morphing systems and structures across scales.
Here, we present a design recipe that transforms an arbitrary plane tiling into a 2D kirigami pattern with a single degree-of-freedom mechanism motion, greatly expanding the known library of  mechanism-based designs. We reveal that  these kirigami patterns, when deformed along their mechanism, have a bulk shape change invariant to the underlying microstructure of the pattern. Experimental observations confirm this unusual kinematic prediction in illustrative classes of designs.  We also exploit this invariance to elicit different elastic responses  in patterns with identical bulk shape change. Finally, we discuss generalizations to  compact and non-planar kirigami, as well as 3D metamaterials, highlighting the broad applicability of our new approach to design. 
\end{abstract}


\maketitle





\section{Introduction}

\begin{figure}[!t]
    \includegraphics[width =1\linewidth]{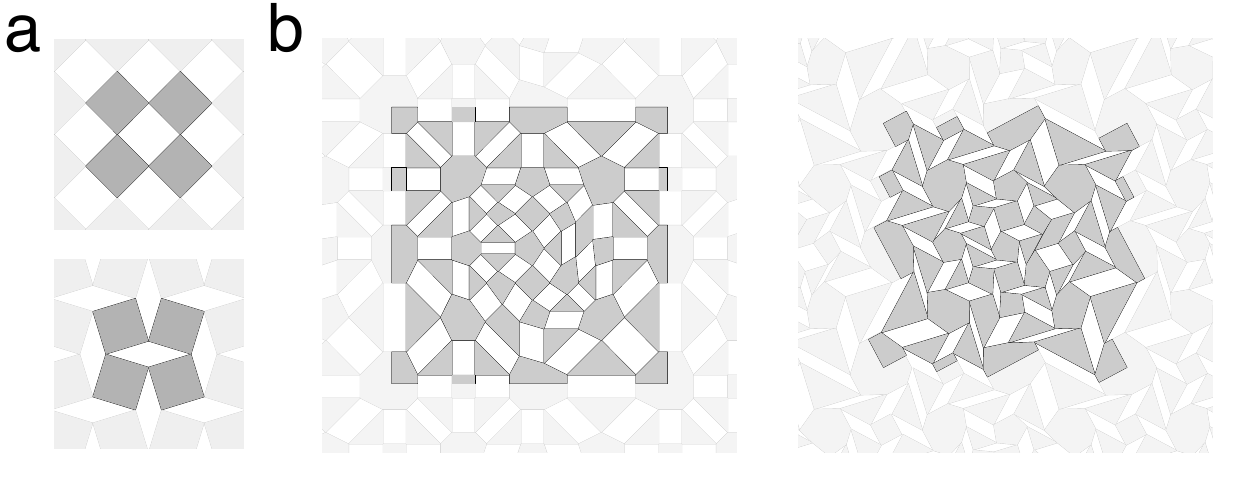}
    \caption{Metamaterials with counter-rotating mechanisms:\;(a) the RS pattern; (b) a generalization constructed from our design recipe. The unit cell of each pattern is emphasized.}
\label{fig:IntroFig}
\end{figure}

Flexible mechanical metamaterials inspired by origami and kirigami  are solids with  unconventional mechanical properties governed primarily by the geometry and topology of their unit cells. A prime example  is the rotating squares (RS)  pattern in Fig.\;\ref{fig:IntroFig}a. Devised  initially as a simple illustration of  auxetic behavior~\cite{grima2000auxetic}, this pattern has since become an archetype for the nonlinear physics of shape-morphing: it is the underlying precursor in the inverse design of complex kirigami shapes \cite{choi2019programming} and targeted deployments \cite{choi2021compact}, exhibits exotic elastic behavior under loads \cite{coulais2018characteristic,czajkowski2022conformal,deng2020characterization,zheng2022continuum,li2025nonlinear}, and can be geometrically tuned to hard-encode bistability in 2D \cite{peng2024programming} and 3D \cite{dang2022theorem}, or “pop-up” buckling \cite{celli2018shape,jin2020kirigami}, or even a desired force-displacement curve \cite{deng2022inverse}. It also represents a common metamaterial platform for hybridized structural systems \cite{dang2024folding,zhao2025modular} and for applications in   mechanical logic  \cite{el2021digital}, soft robotics \cite{jiang2022snakeskin,he2023modular} and biomedical devices \cite{bhullar2014design}.

The RS pattern is popular in science and engineering for two key reasons: it has a  mechanism  \cite{pellegrino1986matrix} or floppy mode \cite{lubensky2015phonons},  that is, this pattern,   when idealized as an assembly
 of rigid panels connected by  perfect pins,  has a single degree-of-freedom (DOF) counter-rotating
 motion at zero energy. This means that, even in  practical engineering settings where the hinges are far from ideal,  it  achieves large overall shape change at little overall stress. Furthermore, its generalizations, obtained by tuning the geometry of the panels and slits, possess novel mechanical properties strongly dictated by this tuning. Another archetype, the Miura origami \cite{koryo1985method}, works in much the same way. It too is endowed with single DOF mechanism motion, has a rich elasticity \cite{wei2013geometric,nassar2022strain, xu2024derivation, feng2025novel}, and is a springboard for generalizations with tunable morphing capabilities \cite{dudte2016, dang2022inverse}. These successes point to the potential of mechanism-based metamaterials, both in revealing new  physics  and in exploiting this physics  in applications. A fundamental
challenge lies in discovering such patterns.

In fact, mechanism-based patterns are rare. Maxwell’s rule \cite{maxwell1864calculation} for counting constraints and DOFs is often not informative, as many patterns appear overconstrained when they do in fact possess mechanisms. Instead, nongeneric symmetries typically encode whether a pattern has a mechanism or not. A major thrust in recent years has been to identify these symmetries in origami and kirigami systems \cite{dieleman2020jigsaw,feng2020designs, choi2021compact, dang2022theorem, dudte2021additive}. This research involves fixing a topology --- the arrangement of panels, slits and/or folds --- and tuning the pattern design to solve for the kinematic compatibility conditions identifying the existence of a mechanism. While not difficult to formulate, this design problem is hard to solve. Most successes have been confined to the simple topologies of familiar motifs, like the RS and  Miura origami patterns.

\begin{figure*}[t!]
    \includegraphics[width =1\linewidth]{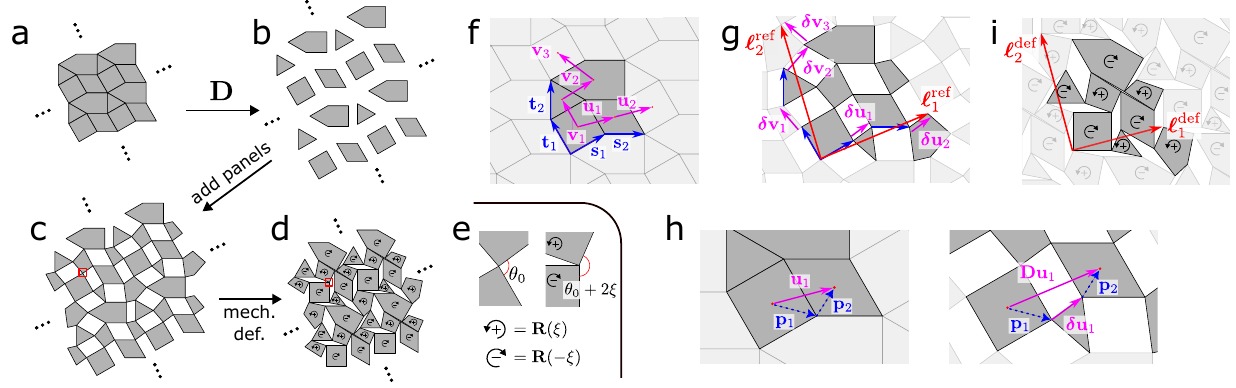}
    \caption{(a-e) Recipe to create 2D mechanism-based metamaterials of arbitrary topology. (a) Start with a plane tiling. (b)
Separate the panels through a tensor $\mathbf{D}$. (c) Add new panels to produce the overall
pattern. (d) The panels counter-rotate
under a mechanism deformation; (e) the slits actuate through angle $\xi$. (f-h) Notation for deriving the effective shape-change of the mechanism motion. (f) Unit cell of the plain tiling and (g) of the metamaterial before deformation. (h) Illustration of how the transformation $\mathbf{D}$ introduces a new set of edge vectors. (i) Unit cell of the metamaterial after deformation.}
\label{fig:DesignPrinciple}
\end{figure*}

In this letter, we reveal a surprising and exceedingly general design recipe for mechanism-based metamaterials.  As highlighted in Fig.\;\ref{fig:IntroFig}b, our designs form striking generalizations of the RS pattern. They are composed of polygonal panels (of a variety of shapes)  and parallelogram slits, arranged into a unit cell of arbitrary size. Each design has a single DOF mechanism. While reminiscent of the network-based approach in Ref.~\cite{acuna2022three} for constructing broad families of perfectly auxetic parallelogram slit metamaterials, our recipe provides quantitative rules for programming  a wide range of targeted   shape-changes. In fact, we unveil that the mechanism often guides the pattern to a bulk (cell-averaged) shape change independent of the underlying microstructure of the panels and slits. We also leverage this invariance to design metamaterials with identical shape change but distinct elastic responses. Finally, we show that our design recipe extends in a natural way to a wide variety of flexible mechanical metamaterials. 







\section{Results and Discussion}

\noindent\textbf{II.1.\,Design recipe ---} We revisit the RS pattern in Fig.\;\ref{fig:IntroFig}a, focusing on what makes it deform as a mechanism. First, note that all slits are parallelograms.
Next, note that the panels counter-rotate about the parallelogram slits to produce the overall mechanism.
In fact, this observation extends to a general family of mechanism-based designs. We show in the Supplemental Material (SM), Sec.\;SM.1.A \cite{suppl} that any 2D pattern of convex panels and parallelogram slits has a single DOF mechanism motion in 2D. While aspects of this result are known to experts \cite{yang2018geometry, singh2021design,acuna2022three, dang2022theorem, zheng2022continuum}, it has almost exclusively been explored in the context of the simple ``checkerboard" topology of the RS pattern. 

As our first main result, we show by way of a three-step recipe that metamaterials with parallelogram slits are ubiquitous, with almost no limitation on their unit cell topology. The key ideas are illustrated in Fig.\;\ref{fig:DesignPrinciple}a-e. Start with a convex polygonal tiling of the plane (Fig.\;\ref{fig:DesignPrinciple}a). Next, translate the panels by mapping their centroids to new points via a 2D homogeneous deformation $\boldsymbol{\varphi}_{\text{des}}(\mathbf{x}) = \mathbf{D} \mathbf{x}$ with  $\mathbf{D} \in \mathbb{R}^{2\times 2}$constrained to have principal stretches larger than one. Such deformations satisfy $\det \mathbf{D} > 1$ and $\min_{\mathbf{e} \in \mathbb{S}^1} |\mathbf{D} \mathbf{e}| > 1,$ and  are guaranteed to separate all the panels (Fig.\;\ref{fig:DesignPrinciple}b). See Sec.\;SM1.B \cite{suppl} for a detailed example of this type of transformation.  Finally, add new polygonal panels by connecting the corner points that were initially coincident in the plane tiling. This recipe yields a pattern with parallelogram slits (Fig.\;\ref{fig:DesignPrinciple}c), that thus has a mechanism (Fig.\;\ref{fig:DesignPrinciple}d). As indicated, its mechanism deformations  counter-rotate the panels by $\mathbf{R}(\pm \xi) \in SO(2)$. Here $\xi$ denotes the \textit{slit actuation} --- it reflects the opening/closing angle of each slit (Fig.\;\ref{fig:DesignPrinciple}e) and  parameterizes the pattern's single DOF mechanism on an interval  $(\xi^{-}, \xi^+)$ where $\xi^{-} \in (-\pi/2, 0)$ and $\xi^+ \in (0, \pi/2)$ depend only on the  design; see Sec.\;SM.1.C \cite{suppl} for more details.

Note that plane tilings with convex polygons are of an uncountable infinity in their variety. Thus, this recipe furnishes a near limitless landscape of designs to explore.  

\vspace*{.5cm}

\noindent\textbf{II.2.\,Effective description of the mechanism ---}  We now quantify the effective shape-change of each pattern's mechanism,ollowing the coarse-graining rule in \cite{zheng2022continuum}. The basic idea  is to   homogenize (or average out) the microstructural rearrangements of the shape-change by exploiting the fact that the plane tiling and the metamaterial, before and after a mechanism deformation, are each described by a repeating unit cell, and thus posses a pair of Bravais lattices vectors that encode the periodicity and overall shape of the cell. The effective description of the mechanism, then, is obtained  by an explicit calculation of the linear transformations that map these Bravais lattice vectors to each other.  To make this idea quantitative, Fig\;\ref{fig:DesignPrinciple}f-i illustrates  the unit cell of the plane tiling and metamaterial, before and after deformation. Overlaid onto these cells are a variety of 2D vectors that form the Bravais lattice vectors. Below, we identify several fundamental geometric relationships between these vectors, which leads to a revealing characterization of the effective shape-change.

Notice first that the edge vectors that form the Bravais lattice vectors of the plane tiling in Fig.\;\ref{fig:DesignPrinciple}f are related to the panel centroid vectors via 
\begin{equation}
    \begin{aligned}\label{eq:latticeVecEquivalents}
        \sum_{i} \mathbf{s}_i = \sum_{i'} \mathbf{u}_{i'}  \quad \text{ and } \quad \sum_{j} \mathbf{t}_j = \sum_{j'} \mathbf{v}_{j'}.
    \end{aligned}
\end{equation}
Transforming this plane tiling to a mechanism-based pattern through $\mathbf{D}$ takes $\mathbf{u}_{i'}$ and $\mathbf{v}_{j'}$ to $\mathbf{D} \mathbf{u}_{i'}$ and $\mathbf{D} \mathbf{v}_{j'}$, respectively. This transformation also yields a new set of edge vectors  $\boldsymbol{\delta} \mathbf{u}_{i'}$ and $\boldsymbol{\delta} \mathbf{v}_{j'}$ (Fig.\;\ref{fig:DesignPrinciple}g) that  satisfy 
\begin{equation}
    \begin{aligned}\label{eq:deltaUiIdents}
        \boldsymbol{\delta} \mathbf{u}_{i'} = (\mathbf{D} - \mathbf{I}) \mathbf{u}_{i'}\quad \text{ and }\quad \boldsymbol{\delta} \mathbf{v}_{j'} = (\mathbf{D} - \mathbf{I}) \mathbf{v}_{j'}
    \end{aligned}
\end{equation}
for elementary geometric reasons (Fig.\;\ref{fig:DesignPrinciple}h).
Summing up these edge vectors, together with their  counterparts $\mathbf{s}_i$ and $\mathbf{t}_j$,  forms a pair Bravais lattice vectors $\boldsymbol{\ell}_1^{\text{ref}}$ and $\boldsymbol{\ell}_2^{\text{ref}}$ that express the periodicity of the metamaterial prior to deformation (Fig.\;\ref{fig:DesignPrinciple}g).  From (\ref{eq:latticeVecEquivalents}) and (\ref{eq:deltaUiIdents}), we relate this periodicity to that of the plane tiling via
\begin{equation}
    \begin{aligned}\label{eq:refBravais}
        \boldsymbol{\ell}_1^{\text{ref}} = \sum_{i} \mathbf{s}_i + \sum_{i'} \boldsymbol{\delta}\mathbf{u}_{i'} = \mathbf{D} \sum_{i'} \mathbf{u}_{i'},\\
        \boldsymbol{\ell}_2^{\text{ref}} = \sum_{j} \mathbf{t}_j + \sum_{j'} \boldsymbol{\delta}\mathbf{v}_{j'} = \mathbf{D} \sum_{j'} \mathbf{v}_{j'}.
    \end{aligned}
\end{equation}

\begin{figure*}[t!]
    \includegraphics[width =0.99\linewidth]{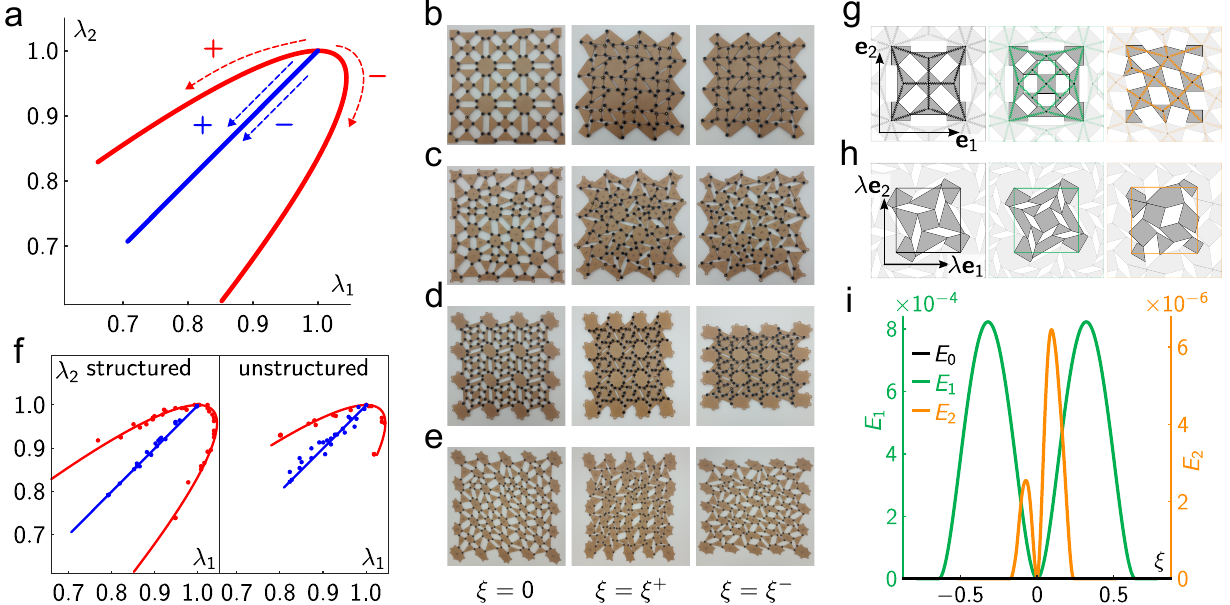}
    \caption{Programming kinematics and elasticity in examples. (a) Effective stretch diagram for conformal kirigami (blue) and neutral Poisson's ratio kirigami (red). (b-c) A structured and unstructured example of conformal kirigami with the same effective shape-change; the undeformed and most actuated states are shown. (d-e) Analogous structured and unstructured examples for neutral Poisson's ratio kirigami. (f) Kinematic validation --- the experimental data markers are overlaid onto the full range of theoretical actuation; the data points are  obtained using the procedure in SM.3.B~\cite{suppl}. (g) Conformal kirigami with identical shape-change and tension springs connecting the panels' centroids. (h) A deformed state  illustrating the bulk shape-change for each example; the unit cell is outlined and the springs are omitted for clarity.  (i) Elastic energy of the metamaterials in (g) as a function of the actuation angle. }
\label{fig:Example}
\end{figure*}

After deformation, the metamaterial changes shape but remains periodic, furnishing the deformed  Bravais lattice vectors $\boldsymbol{\ell}_1^{\text{def}}$ and $\boldsymbol{\ell}_2^{\text{def}}$ (Fig.\;\ref{fig:DesignPrinciple}i). As each $\mathbf{s}_i$ and $\mathbf{t}_j$ rotates by $\mathbf{R}(-\xi)$ and each $\boldsymbol{\delta} \mathbf{u}_{i'}$ and $\boldsymbol{\delta}\mathbf{v}_{j'}$ counter-rotates by $\mathbf{R}(\xi)$, these lattice vectors satisfy   
\begin{equation}
    \begin{aligned}\label{eq:defBravais}
       \boldsymbol{\ell}_1^{\text{def}} &= \sum_{i} \mathbf{R}(-\xi) \mathbf{s}_i + \sum_{i'} \mathbf{R}(\xi) \boldsymbol{\delta}\mathbf{u}_{i'} \\
       &= \Big(\mathbf{R}(-\xi) + \mathbf{R}(\xi) ( \mathbf{D} - \mathbf{I})\Big) \sum_{i'} \mathbf{u}_{i'}, \\
       \boldsymbol{\ell}_2^{\text{def}} &= \sum_{j} \mathbf{R}(-\xi) \mathbf{t}_j + \sum_{j'} \mathbf{R}(\xi) \boldsymbol{\delta}\mathbf{v}_{j'} \\
       &= \Big(\mathbf{R}(-\xi) + \mathbf{R}(\xi) ( \mathbf{D} - \mathbf{I})\Big) \sum_{j'} \mathbf{v}_{j'},
    \end{aligned}
\end{equation}
using the identities in (\ref{eq:latticeVecEquivalents}) and (\ref{eq:deltaUiIdents}).

The shape-change induced by the mechanism deformation can be cataloged by a tensor  $\mathbf{A}_{\text{eff}} \in \mathbb{R}^{2\times2}$  that maps the reference Bravais lattice vectors of the metamaterial to the deformed ones  through the formulas   $\mathbf{A}_{\text{eff}} \boldsymbol{\ell}_{i}^{\text{ref}} = \boldsymbol{\ell}_{i}^{\text{def}}$ for $i =1,2$. As a central finding,  we obtain
\begin{equation}
    \begin{aligned}\label{eq:Feff}
        \mathbf{A}_{\text{eff}}(\xi) = (\mathbf{R}(-\xi) - \mathbf{R}(\xi)) \mathbf{D}^{-1} + \mathbf{R}(\xi) 
    \end{aligned}
\end{equation}
after pre-multiplying the equations in (\ref{eq:refBravais}) by $\mathbf{D}^{-1}$ and substituting the resulting formulas into (\ref{eq:defBravais}). We call $\mathbf{A}_{\text{eff}}$ the \textit{shape tensor} (see also \cite{zheng2022continuum}).  As emphasized, this tensor evolves kinematically with the slit actuation $\xi$, quantifying how the  mechanism deforms the overall pattern. Note that, although we leveraged periodicity to obtain this characterization, the shape tensor  $\mathbf{A}_{\text{eff}}$ derived here can also be used to quantify  the effective shape-change of metamaterials generated from   an aperiodic or unstructured plane tiling using our design recipe; see Sec.\;SM1.D \cite{suppl} for the derivation.

\vspace*{.5cm}
\noindent \textbf{II.3.\,Examples ---} Equation (\ref{eq:Feff}) has  surprising implications for design. Notice that the shape tensor depends on the geometric parameters of the pattern only through the tensor $\mathbf{D}$ but is  agnostic to the underlying plane tiling used to generate it. In other words,  the pattern's shape change is independent of  its microstructure. This surprising feature also offers practical utility for engineering design. In particular, by tuning $\mathbf{D}$, we can program metamaterial designs with a myriad of distinct effective shape-changes including   axial shape-change, pure shear or a mixture of both.

As a simple set of illustrations, we consider examples with axial effective shape-change, which are obtained by constraining the shape tensor to be diagonal  $\mathbf{A}_{\text{eff}}(\xi)   = \sum_{i =1,2} \lambda_i(\xi) \mathbf{e}_i \otimes \mathbf{e}_i$. In Sec.\;SM.2.A~\cite{suppl}, we show that this setting requires that $\mathbf{D}^{-1} =   \frac{1}{2} \mathbf{I} + \mathbf{R}(\frac{\pi}{2}) \sum_{i =1,2} d_i \mathbf{e}_i \otimes \mathbf{e}_i$ for some $d_1,d_2 \in \mathbb{R}^2$, which furnishes the stretches
\begin{equation}
    \begin{aligned}
        &\lambda_i(\xi) =\cos \xi + 2 d_i \sin \xi
    \end{aligned}
\end{equation}
for $i = 1,2$. Note that $d_1, d_2$ are subject to  inequality constraints, detailed in Sec.\;SM.2.B~\cite{suppl}, which ensure that $\mathbf{D}$ yields a physical pattern. Besides the stretches, a typical quantity of interest is  the Poisson's ratio, signifying whether the pattern is auxetic or not.  This parameter is defined for any diagonal shape tensor with  actuation $\xi$ as  $\nu(\xi) = \lambda_2'(\xi) \lambda_1(\xi)/\big(\lambda_1'(\xi) \lambda_2(\xi)\big)$ (see Sec.\;SM.2.C~\cite{suppl}).  

We emphasize two special families  of these kirigami designs in Fig.\;\ref{fig:Example}. The first is a case of \textit{conformal kirigami}, obtained by setting $d_1 = d_2 = 0$. These patterns have stretch $\lambda_{1}(\xi) = \lambda_2(\xi) = \cos \xi$, corresponding to the straight blue line with slope $1$ in the stretch diagram in Fig.\;\ref{fig:Example}a. The second  are patterns with  \textit{neutral Poisson's ratio}, obtained by setting $d_2 = 0$ and $d_1 \neq 0$. Their kinematics  are shown with the red line in Fig.\;\ref{fig:Example}a. Notice that the undeformed pattern contracts in the $\mathbf{e}_2$-direction regardless of whether it is  stretched or compressed in the $\mathbf{e}_1$-direction.  This unusual actuation furnishes  a Poisson's ratio $\nu(\xi) \approx -\frac{1}{2d_1} \xi$ when $|\xi| \ll 1$, that thus  switches the pattern from being auxetic to not auxetic at $\xi =0$. 

By considering different plane tilings as input to our design recipe, each of these classes constitutes an enormous family of metamaterial designs programmed with the same shape tensor $\mathbf{A}_{\text{eff}}$. We fabricate  and test two designs in each class to validate that their effective shape-change is the same,  as predicted by $\mathbf{A}_{\text{eff}}$, and thus independent of microstructure. One design is structured  and composed of a $3\times3$ array of cells; the other is unstructured and quasi-random (see Sec.\;SM.2.D~\cite{suppl}). Each pattern is made of acrylic panels, connected by circular pins that allow free rotation in the plane. We choose this specific fabrication strategy since pin-joints best reflect the ideal mechanism behavior described by the theory. Replacing these joints with flexures can  yield a myriad of other inhomogeneous planar and non-planar soft modes of deformation, for which the shape tensor $\mathbf{A}_{\text{eff}}$ is predicted to characterize the shape-changing  response of the pattern in a local (though not necessarily global) sense~\cite{zheng2023modelling}.

Fig.\;\ref{fig:Example}b-e shows all four of the fabricated samples in their undeformed ($\xi = 0$) and most actuated  ($\xi= \xi^{\pm}_{\text{exp}}$) states; Fig.\;\ref{fig:Example}f compares their kinematics to that of the theory. Overall, the agreement  is excellent --- the structured and unstructured examples in both cases exhibit the same overall shape-change,  in line  with the theory, even though their microstructures  are completely different. Notably, however,  the  range of actuation is muted  for the structured examples.   We show in Sec.\;SM.3.C~\cite{suppl}  that this discrepancy is entirely due to the  pins, as  they  prevent the slits from fully closing.

Going beyond kinematics, we can exploit the decoupling of overall    shape  and microstructure to produce novel elastic behavior. We focus on a toy model of elasticity here to highlight this point, leaving it to future work to apply these ideas more broadly. Fig.\;\ref{fig:Example}g   considers  three distinct examples of  conformal kirigami. Each  has a unit cell that repeats along  $\mathbf{e}_1$ and $\mathbf{e}_2$, as shown, but with  fundamentally different arrangements  of   panels and slits and thus distinct microstructures. Each is also programmed to contract by the same conformal effective shape-change $\lambda(\xi) = \cos \xi$ under a mechanism deformation (Fig.\;\ref{fig:Example}h). However, the  elastic behavior of these patterns during this shape-change differs dramatically when we  attach  ``tensile" springs (i.e., rubber bands) between the centroids of adjacent panels. In this setting,  the   actuation energy of   a  unit cell with $N$ such springs  is 
\begin{align}\label{eq:elastic}
    E(\xi) =   \sum_{i=1}^N \big(\max\{ l_i(\xi) - l_i^{\text{ref}}, 0\} \big)^2
\end{align}
where $l_i^{\text{ref}}$  and $l_i(\xi)$ are the undeformed and actuated length of the $i^{\text{th}}$ spring in the cell, respectively.  Note that the springs  compress freely due to the maximizaton.

\begin{figure*}[t!]
    \includegraphics[width =0.9\linewidth]{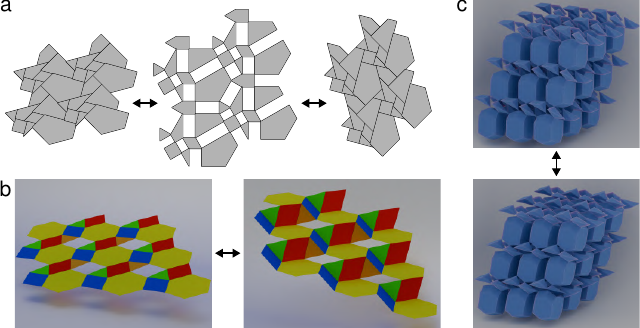}
    \caption{Generalizations. (a) Compact 2D kirigami has an open configuration composed of rectangular slits and a mechanism that evolves it to  two fully-closed states. (b) Non-planar kirigami with a single DOF mechanism can be obtained from our recipe by replacing some parallelogram slits with panel. (c) 3D generalization of our design recipe --- the metamaterial shown has a 3 DOF mechanism.} 
\label{fig:Generalizations}
\end{figure*}

 Fig.\;\ref{fig:Example}h plots this  energy for the three examples. In the first example (black), the pattern starts mirror-symmetric with the springs overlapping the  touching points of adjacent panels. Each spring then compresses under the counter-rotating mechanism actuation, yielding no elastic resistance. The second example (green) is  also a mirror-symmetric. However, some of its springs undergo significant tension during actuation even though the overall pattern contracts; this makes the pattern elastically tristable with a symmetric energy barrier to jump from the $\xi = 0$ to the $\xi = \xi^{\pm}$ states. The final example (orange) is not mirror-symmetric --- the springs are more activated  for $\xi >0$  compared to $\xi <0$ actuation,  leading to an asymmetric energy barrier. Interestingly this energy barrier is two orders of magnitude lower than the second example, even though the springs in both cases have the same unit modulus in (\ref{eq:elastic}). All told,   this simple illustration shows  how microstructural differences provide a pathway for elastic tuning, even in patterns with the same overall shape-change.   


\vspace*{.5cm}

\noindent\textbf{II.4.\;Generalizations ---} We end by  discussing  generalizations of our mechanism-based designs. These ideas are supported by  analysis in the supplmental~\cite{suppl}, but are by no means  fully developed. We hope they serve to showcase the broad applicability of our design strategy and inspire further research in this direction.

Recall that we used a tensor $\mathbf{D}$  to uniformly separate the plane tiling and generate the parallelogram slit design. While this  choice lead to a compelling description of the effective shape-change, we can also separate the panels non-uniformly  while keeping the rest of   design recipe  in Fig.\;\ref{fig:DesignPrinciple}a-d the same. This extra freedom allows us to generate  designs with rectangular slits (see Sec.\;SM.4 for details~\cite{suppl}). As shown in Fig.\;\ref{fig:Generalizations}a, these designs are  \textit{compact kirigami patterns}, akin to those in \cite{choi2021compact}, but without restrictions on cell topology  ---  all of the slits  fully close when the mechanism is actuated  to its  limit ($\xi^{\pm} = \pm \frac{\pi}{4}$).  

We can also obtain non-planar mechanism-based kirigami through another  modification of the recipe. All 2D parallelogram slit designs  become much floppier in 3D; in fact, they become isostatic in  a counting sense in that their  3D floppy modes scale   $\sim N$ for a pattern with $N \times N$  cells  (see Sec.\;SM.5.A~\cite{suppl}). To reach a finite but non-zero number of mechanisms as $N \rightarrow \infty$, our fix is to replace exactly one or two parallelogram slits in each unit cell with panels, letting the common edges between panels become fold lines. As explained in Sec.\;SM.5.B~\cite{suppl} (using \cite{feng2020designs,foschi2022explicit}), this replacement rigidifies the unit cell but does not fully eliminate the 3D generalizations of the 2D bulk mechanism.   Indeed, Fig.\;\ref{fig:Generalizations}b shows an example where two slits per cell are replaced with panels to obtain a mechanism-based design with exactly one DOF. 

Lastly, we note that the design recipe in Fig.\;\ref{fig:DesignPrinciple}a-d
has a direct generalization to 3D. Simply start with a space tiling of convex polyhedra, rather than a plane tiling, and replace a 2D tensor $\mathbf{D}$ with a 3D one satisfying $\det \mathbf{D} > 1$ and $\min_{\mathbf{e} \in \mathbb{S}^2} |\mathbf{D} \mathbf{e}| > 1$ to separate  the polyhedra. Fig.\;\ref{fig:Generalizations}c shows one such design. The mechanism kinematics of this 3D metamaterial is more nuanced then its 2D counterparts.  As detailed in  Sec.\;SM.6~\cite{suppl}, it  has a 3 DOF family of mechanism deformations parameterized by a single rotation $\mathbf{R} \in SO(3)$,  the natural extension of the 2D counter-rotating  mode;  it  also has a novel infinitesimal twisting mode with no 2D analogue, though curiously this mode does not extend to a fully nonlinear mechanism at large deformation. Interestingly, the    same  reasoning as (\ref{eq:latticeVecEquivalents}-\ref{eq:Feff}) implies that  the 3D shape tensor here evolves with the mechanism actuation $\mathbf{R}$ via $\mathbf{A}_{\text{eff}} = (\mathbf{I} - \mathbf{R}) \mathbf{D}^{-1} + \mathbf{R}$. Consequently, the effective shape-change of these designs is again independent of microstructure. 

\section{Conclusions}
This letter revealed a surprising and simple strategy for designing mechanism-based metamaterials. In contrast with the typical  design paradigm, where certain canonical unit cells  are geometrically tuned to achieve new designs, we  leveraged  the  inherent compatibility of  parallelogram slits and mechanisms  to construct designs with  arbitrarily large unit cells  and striking complexity.   
We then used theory and experiments to show that these patterns, when deformed along their mechanism, possess a cell-averaged shape change  independent of their microstructure. We also exploited this invariance to program different elastic responses  in patterns with the same bulk shape-change, and  showed how the design strategy naturally generalizes to a variety of  metamaterial platforms in both two and three dimensions. All told, our  work here greatly expands the known library of mechanism-based designs. We envision that the decoupling of microstructure and shape-change intrinsic to our designs can perhaps be used to functionalize these metamaterials in novel ways for applications in soft robotics, biomedical devices, and deployable structures. 
One intriguing  functionalization is in the realm of active phased-array antennas \cite{jiang2025abnormal}, where dipoles serve as programmable ``atoms” attached to  each panel. Reconfigurable metasurfaces of nearly arbitrary topology provide a highly  tunable platform for  adaptive, real-time beam steering and wave shaping through  precise control  of  orientation and periodicity within a single structure.

\begin{acknowledgments}
 YP and PP acknowledge support from   NSF (CMMI-CAREER-2237243) and ARO (ARO-W911NF2310137). AR and PC acknowledge support from NSF (CMMI-2045191).
\end{acknowledgments}

\bibliographystyle{apsrev4-1}
\bibliography{PRLRef1}


\end{document}


\title{Supporting Information for ``Mechanism-based metamaterials with microstructurally invariant shape-change''}

\author{Yingchao Peng}
\affiliation{Aerospace and Mechanical Engineering, University of Southern California, Los Angeles, CA 90014, USA}
\author{Asifur Rahman}%
\affiliation{Civil Engineering, Stony Brook University, Stony Brook, NY 11794, USA}
\author{Paolo Celli}
\affiliation{Civil Engineering, Stony Brook University, Stony Brook, NY 11794, USA}
\author{Paul Plucinsky}
\email{plucinsk@usc.edu}
\affiliation{Aerospace and Mechanical Engineering, University of Southern California, Los Angeles, CA 90014, USA}




\maketitle

\section{Parallelogram slits and mechanisms}

In this paper, we introduce a broad class of mechanism-based metamaterials composed of convex polygonal panels and parallelogram slits. Here we explain why these patterns have a single DOF mechanism, and provide a quantitative description of their kinematics.

\subsection{Characterization of the mechanism kinematics}

\begin{figure}[h!]
    \centering
    \includegraphics[width=0.5\linewidth]{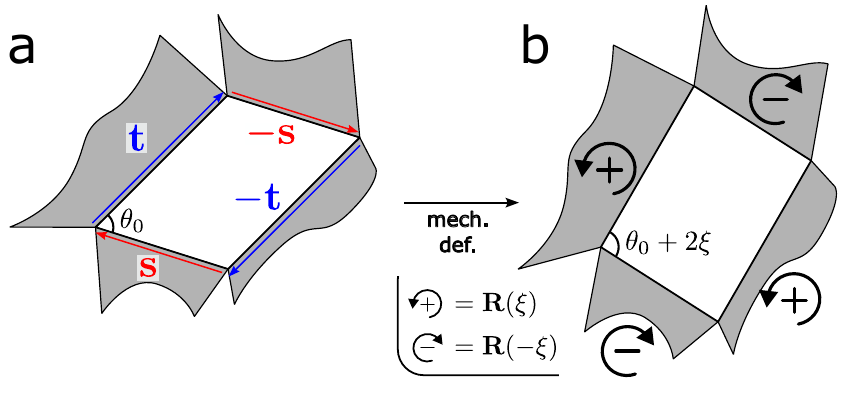}
    \caption{\textit{Mechanism kinematics of a single parallelogram slit. (a) The slit is parameterized by edge vectors $\pm \mathbf{s}$ and $\pm \mathbf{t}$ and corresponding sector angles  $\theta_0$ and $\pi - \theta_0$. (b)  Under a mechanism deformation, the panels counter-rotate by  $\mathbf{R}(\pm \xi) \in SO(2)$ and $\theta_0$ goes to  $\theta _0+ 2 \xi$.}  }
    \label{fig:SingleSlit}
\end{figure}

\noindent \textbf{Local mechanism kinematics.} Let's begin by studying a single  parallelogram slit (Fig.\;\ref{fig:SingleSlit}a), where   $\pm \mathbf{s}$ and $\pm \mathbf{t} \in \mathbb{R}^2$ parameterize the edges of the slit with $-\mathbf{s} \cdot \mathbf{t} = |\mathbf{s}| |\mathbf{t}|\cos \theta_0$. The panels in this setup are arbitrary convex polygons. Deforming this  pattern as a mechanism requires rotating the four panels in such a way that the slit's closed loop (i.e., the fact that $\mathbf{s} + \mathbf{t} - \mathbf{s} - \mathbf{t} = \mathbf{0})$ remains a closed loop after deformation. Let the panel indicated by the vector  $\mathbf{s}, \mathbf{t}, -\mathbf{s}, -\mathbf{t}$ rotate by $\mathbf{R}_1$, $\mathbf{R}_2$, $\mathbf{R}_3$, $\mathbf{R}_4 \in SO(2)$, respectively. Then, closing the loop on deformation requires  
\begin{equation}
\begin{aligned}\label{eq:loopCompat}
    (\mathbf{R}_1 - \mathbf{R}_3)\mathbf{s} + (\mathbf{R}_2 - \mathbf{R}_4) \mathbf{t} =  \mathbf{0}.
\end{aligned}
\end{equation}
Writing $\mathbf{R}_1 = \mathbf{R}_3 = \mathbf{R}(-\xi)$ and $\mathbf{R}_2 = \mathbf{R}_4 = \mathbf{R}(\xi)$ for 
\begin{equation}\label{eq:Rxi}
    \begin{aligned}
        \mathbf{R}(\xi) = \begin{pmatrix} \cos \xi & -\sin \xi \\ \sin \xi & \cos \xi \end{pmatrix}
    \end{aligned}
\end{equation}
produces a family of solutions to this compatibility condition, parameterized by $\xi \in (-\frac{1}{2} \theta_0, \frac{1}{2} ( \pi - \theta_0)\big)$  where $-\frac{1}{2} \theta_0 < 0 < \frac{1}{2} ( \pi - \theta_0)$ indicates the two values of $\xi \in (-\frac{\pi}{2}, \frac{\pi}{2})$ corresponding to a fully closed slit. One such mechanism deformation is shown in Fig.\;\ref{fig:SingleSlit}b.

\vspace*{.2cm}
\noindent \textbf{Uniqueness.} The mechanism of this single slit is fully characterized by this counter-rotating family of deformations. To see this, we use an argument based on linearization and the fact that the number of infinitesimal mechanisms of a pattern is greater than or equal to the corresponding number of nonlinear mechanisms. The general idea is this:  we have a parameterization for a single DOF mechanism, and mechanisms are by definition continuous motions. So all we need to do to demonstrate uniqueness is to perturb along this parameterization and show that the number of infinitesimal mechanisms is always one, i.e., the same as the dimension of our mechanism parameterization. 

\begin{figure}[b!]
    \centering
    \includegraphics[width=0.65\linewidth]{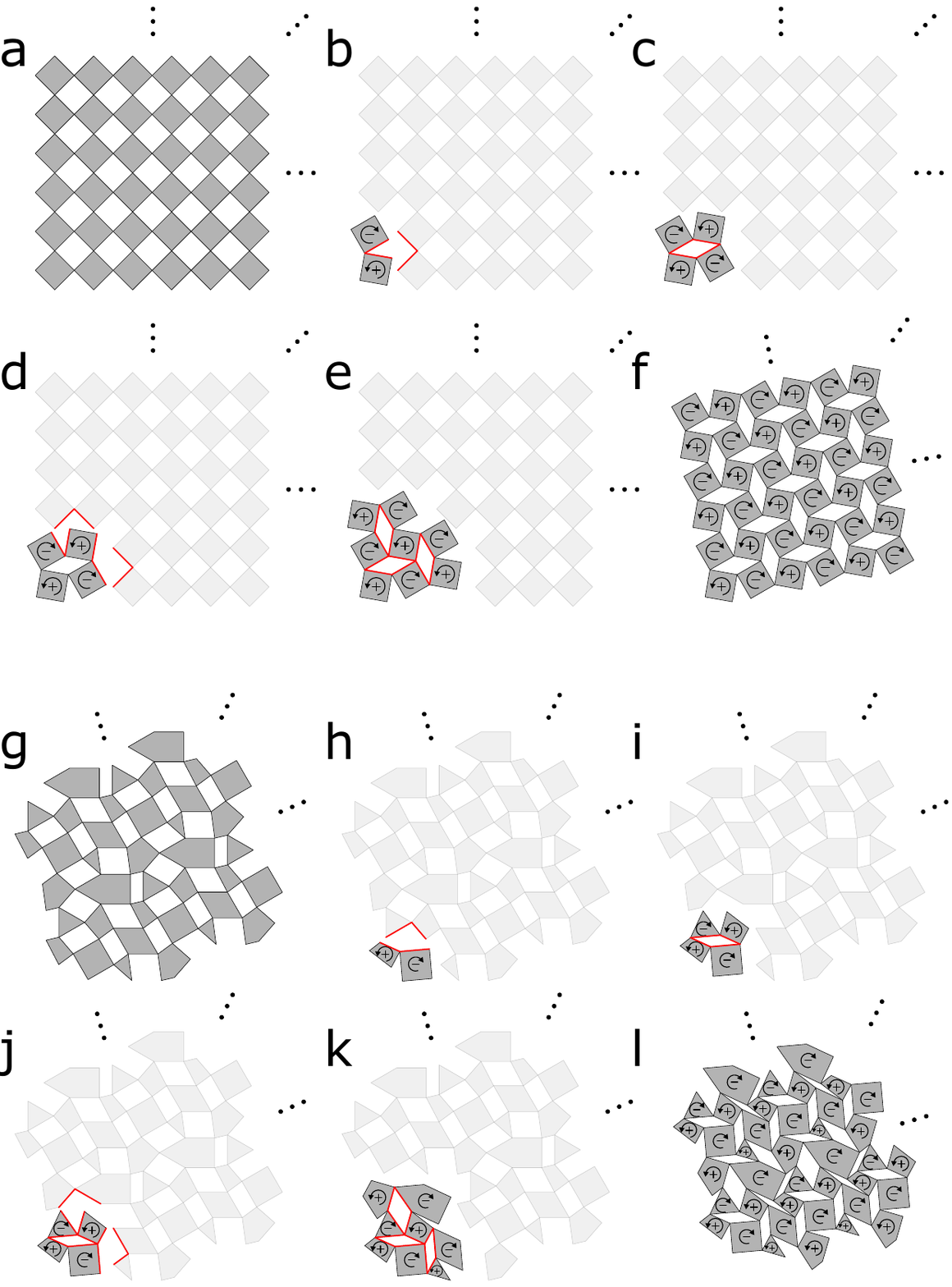}
    \caption{\textit{The mechanism kinematics of global patterns with convex polygons and parallelogram slits. (a-f) The example of the RS pattern: Counter-rotating just two panels implies by iteration that the  other panels must counter-rotate to achieve a compatible mechanism deformation. (g-l) An example of a different topology; the iterative procedure works just  the same. }} 
    \label{fig:ParaSlitsConstruct}
\end{figure}

Substitute $\mathbf{R}_1 = \mathbf{R}(-\xi + \delta \phi_1), \mathbf{R}_2 = \mathbf{R}(\xi + \delta \phi_2)$, $\mathbf{R}_3 = \mathbf{R}(-\xi + \delta \phi_3), \mathbf{R}_4 = \mathbf{R}(\xi + \delta \phi_4)$ into (\ref{eq:loopCompat}) to obtain 
\begin{equation}
    \begin{aligned}\label{eq:compatLin}
        &\big(\mathbf{R}(-\xi + \delta \phi_1) - \mathbf{R}(-\xi + \delta \phi_3)\big) \mathbf{s} + \big(\mathbf{R}(\xi + \delta \phi_2) - \mathbf{R}(\xi + \delta \phi_4) \big) \mathbf{t}  \\
        &\qquad =  (\delta \phi_1 - \delta \phi_3) \mathbf{W} \mathbf{R}(-\xi) \mathbf{s} + (\delta \phi_2 - \delta \phi_4) \mathbf{W} \mathbf{R}(\xi) \mathbf{t} + O(|(\delta \phi_1, \delta \phi_2, \delta \phi_3, \delta \phi_4)|^2), 
    \end{aligned}
\end{equation}
where $\mathbf{W} = \mathbf{R}(\tfrac{\pi}{2})$. Note that $ \mathbf{R}(-\xi) \mathbf{s}$ and $\mathbf{R}(\xi) \mathbf{t}$ are linearly independent for all $\xi \in (-\frac{1}{2} \theta_0, \frac{1}{2} ( \pi - \theta_0)\big)$. Thus, (\ref{eq:compatLin}) vanishes at leading order if and only if $\delta \phi_1 = \delta \phi_3 = \delta \gamma - \delta \xi$ and $\delta \phi_2 = \delta \phi_4 = \delta \gamma + \delta \xi$ for some $\delta \gamma, \delta \xi \in \mathbb{R}$. Here, $\delta \gamma$ is an infinitesimal rigid motion of the entire pattern, while $\delta \xi$ is a linear perturbation of non-linear parameterization of the counter-rotating family of mechanism deformation. So there is exactly one infinitesimal mechanism, excluding the trivial rigid motion. This proves that our $\xi$-parameterized family of mechanism deformations is exhaustive. 

\vspace*{.2cm}

\noindent \textbf{Global mechanism kinematics.} We now show that patterns composed solely of convex polygonal panels and parallelogram slits have a single DOF mechanism. The argument is  iterative: Consider first the RS pattern in Fig.\;\ref{fig:ParaSlitsConstruct}a, and begin by rotating the two lower left panels relative to each to initialize a mechanism motion (Fig.\;\ref{fig:ParaSlitsConstruct}b). This deformation breaks the pattern up at the slit initially joining the two  the panels (in red). To reconnect the slit, we must rotate the two neighboring panels in an alternating fashion (Fig.\;\ref{fig:ParaSlitsConstruct}c), in accordance with the single-slit characterization. Doing so, however, results in more broken slits (Fig.\;\ref{fig:ParaSlitsConstruct}d).  To correct these, we simply iterate on our single slit characterization (Fig.\;\ref{fig:ParaSlitsConstruct}e) until the entire pattern has been deformed (Fig.\;\ref{fig:ParaSlitsConstruct}f).  

Importantly, this iterative procedure does not depend on the pattern topology. Fig.\;\ref{fig:ParaSlitsConstruct}g-l shows, for instance, the same  procedure applied to the example  with triangles, quads and pentagons from the design recipe in Fig.\;2 of the main text. The key point is that each parallelogram slit demands that panels on   opposite sides rotate the same. This local condition propagates from slit to slit, leading to a global characterization of the mechanism where the panels alternate between $``+"$  and $``-"$ rotations, respectively.

\subsection{Generating a parallelgoram slit design}

\begin{figure}[h!]
    \centering
    \includegraphics[width=0.7\linewidth]{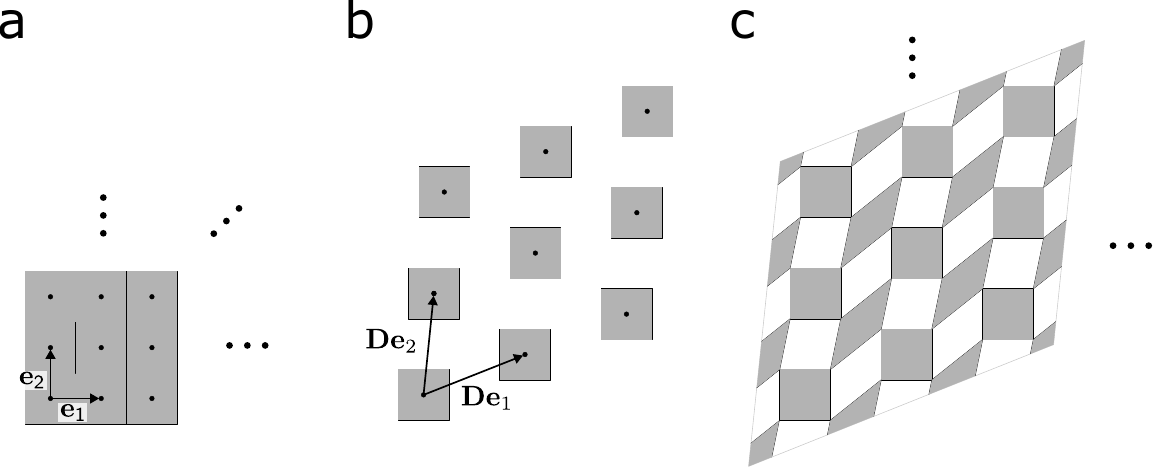}
    \caption{\textit{Illustration of the separation tensor. (a) Square tiling of the plane with centroids indicated. (b) Application of the separation tensor $\mathbf{D}$, with the separated centroids highlighted by the vectors $\mathbf{D}\mathbf{e}_1$ and $\mathbf{D}\mathbf{e}_2$. (c) Panels added using our design recipe to form a parallelogram slit pattern.}}
    \label{fig:separationtensor}
\end{figure}

Here we provide a concrete example of  how the separation tensor $\mathbf{D} \in \mathbb{R}^{2\times2}$, introduced in the main text, is used to generate a parallelogram slit design from a plane tiling.  Consider the unit square tiling of the plane in Fig.\;\ref{fig:separationtensor}a.  Notice that  the centroid of each square is given by an element of the  $\mathbb{Z}^2$ lattice, i.e., $(i,j) \in \mathbb{Z}^2$ where $i$ and $j$ are integers.  To  transform this tiling into a metematerial, we first deform these centroids via 
\begin{align}
\begin{pmatrix}  i \\  j \end{pmatrix} \mapsto \mathbf{D} \begin{pmatrix} i \\ j  \end{pmatrix} \quad \text{ for all } (i,j) \in \mathbb{Z}^2,
\end{align}
 but preserve each panels shape, size  and orientation around the  deformed centroid.  This transformation is shown in Fig.\;\ref{fig:separationtensor}b. Notice that the Bravais lattice vectors of the unit square tiling $\mathbf{e}_1$ and $\mathbf{e}_2$ are deformed to $\mathbf{D}\mathbf{e}_1$ and $\mathbf{D} \mathbf{e}_2$. These vectors then form the Bravais lattices vectors, defining the periodicity of the pattern in Fig.\;\ref{fig:separationtensor}c, obtained  by connecting the initially coincident corner points by  a panel. The assumptions $\det \mathbf{D} > 1$ and $\min_{\mathbf{e} \in \mathbb{S}^1} |\mathbf{D} \mathbf{e}| >1$ guarantee that the transformation separates all the panels in Fig.\;\ref{fig:separationtensor}b, leading to a design in Fig.\;\ref{fig:separationtensor}c composed of convex polygonal panels and parallelogram slits.

\subsection{The limit states of the mechanism}

\begin{figure}[h!]
    \centering
    \includegraphics[width=0.8\linewidth]{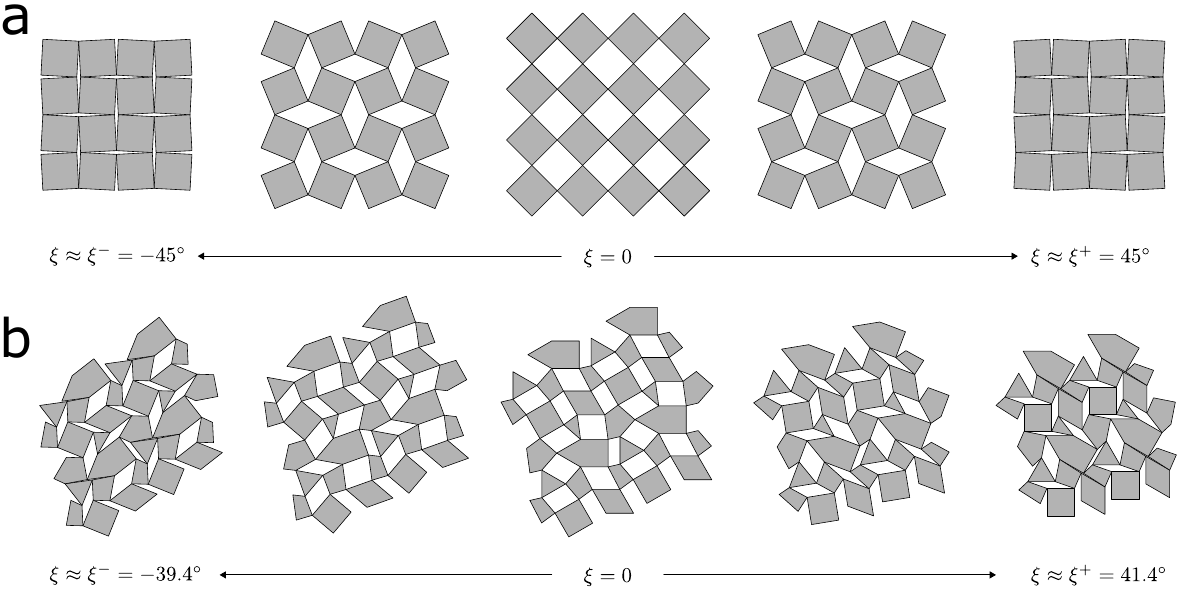}
    \caption{\textit{Illustration of the mechanism motion for (a) the RS pattern  and (b) an example composed of triangles, quads, and pentagon panels. The bounds $\xi^{\pm}$ represent the first negative and positive $\xi$ for which (at least) one slit in the pattern closes.}  }
    \label{fig:MechDefBounds}
\end{figure}

 Having demonstrated the existence and uniqueness of the single DOF mechanisms in our setup,  we now briefly discuss the interval  $\xi \in (\xi^-, \xi^+)$ parameterizing the actuation.   Fig.\;\ref{fig:MechDefBounds}a shows the mechanism evolution for the RS pattern; Fig.\;\ref{fig:MechDefBounds}b does likewise for the pattern introduced in Fig.\;2 of the main text. Notice that the shape of the $i = 1,2,3,\ldots$ slit in the reference configuration ($\xi = 0$) is described by a sector angle  $\theta_i \in (0,\pi)$ for which $\theta_i \rightarrow \theta_i + 2 \xi$ on actuation (recall $\theta_0$ in Fig.\;\ref{fig:SingleSlit}). This slit then closes when $\xi = - \frac{1}{2} \theta_i$ or $\xi = \frac{1}{2} ( \pi - \theta_i)$. Thus, we define $\xi^{-}$ and $\xi^+$ for a given pattern as the first instance of slit closure for negative and positive $\xi$, respectively, i.e., 
\begin{equation}
\begin{aligned}\label{eq:xiPM}
\xi^- = \max_{\text{all }  i = 1,2,3\ldots \text{ slits}} -\frac{1}{2} \theta_i, \quad \xi^{+} = \min_{\text{all } i = 1,2,3\ldots \text{ slits}} \frac{1}{2} ( \pi - \theta_i).
\end{aligned}
\end{equation}

The interval $(\xi^{-}, \xi^+)$ depends on the pattern design. For the RS pattern in Fig.\;\ref{fig:MechDefBounds}a, all the slits are squares in the reference state, which means from (\ref{eq:xiPM}) that $\xi^{\pm} = \pm 45^{o}$. Note that every slit  closes at these two bounds simultaneously.  For the second example in Fig.\;\ref{fig:MechDefBounds}b, there are five distinct types of slits. One  closes first at a negative $\xi$ value of $\xi^{-} = -39.4^{o}$. A different one closes first at a positive $\xi$ value of $\xi^{+} = 41.4^{o}$.

\subsection{Effective shape-change for aperiodic and unstructured metamaterial designs}

\begin{figure}
    \centering
    \includegraphics[width=0.8\linewidth]{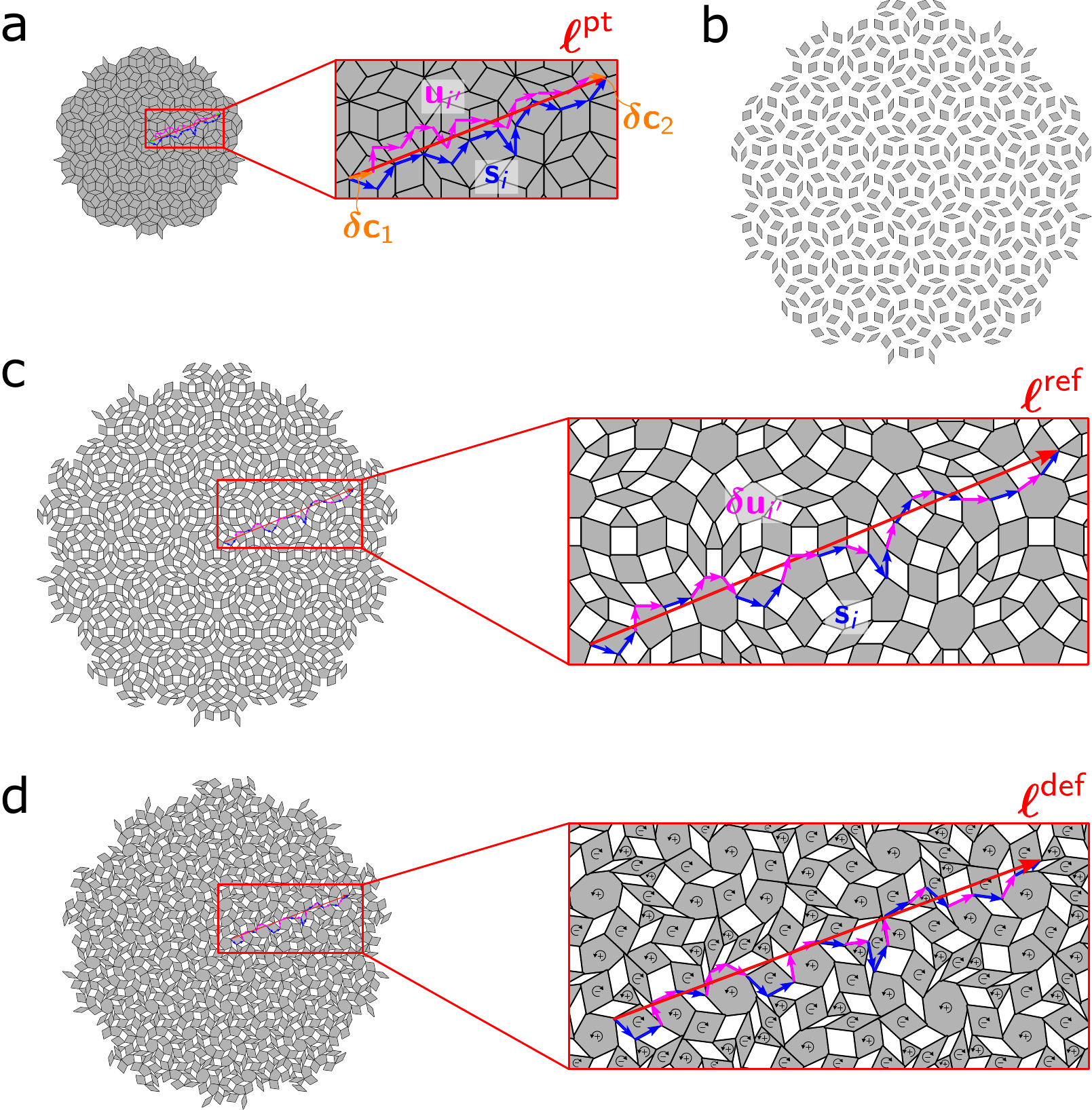}
    \caption{\textit{Illustration of the design recipe and effective shape-change for aperiodic tiling (Penrose tiling). (a) Classical Penrose tiling, with a large sum of edge vectors $\boldsymbol{\ell}^{\emph{pt}}$ highlighted. (b) The pattern after applying the separation tensor $\mathbf{D}$. (c) The resulting aperiodic parallelogram slit pattern with the sum of edge vector $\boldsymbol{\ell}^{\emph{ref}}$ highlighted. (d) A mechanism deformation of this pattern, with $\boldsymbol{\ell}^{\emph{def}}$ denoting the distortion of $\boldsymbol{\ell}^{\emph{ref}}$.}}
    \label{fig:penrose}
\end{figure}

The main text focused on a characterization of the effective shape-change of periodic designs. Our design recipe can also produce aperiodic or unstructured metamaterials that have parallelogram slits and a single DOF mechanism by starting from a corresponding aperiodic or unstructured  plane tiling. Here, we  show that the effective shape-change of such metamaterials is also characterized in an approximate sense by the shape tensor $\mathbf{A}_{\text{eff}}(\xi)$. We use a metamaterial constructed from the aperiodic Penrose tiling in Fig.\;\ref{fig:penrose} to illustrate the derivation. The  essence of the characterization is this: after transforming the plane tiling (Fig.\;\ref{fig:penrose}a) through a separation tensor $\mathbf{D}$ (Fig.\;\ref{fig:penrose}b), we obtain a  metamaterial (Fig\;\ref{fig:penrose}c) that has a counter-rotating single DOF mechanism (Fig.\;\ref{fig:penrose}d). The sum of any $M$ connected   edge vectors in the reference metamaterial $\boldsymbol{\ell}^{\text{ref}}(M)$ distorts under a mechanism deformation to $\boldsymbol{\ell}^{\text{def}}(M)$ (see the right of Fig.\;\ref{fig:penrose}c-d). Below, we prove that, for any such sum, 
\begin{align}\label{eq:MellRefellDef}
\frac{|\mathbf{A}_{\text{eff}}(\xi) \boldsymbol{\ell}^{\text{ref}}(M) - \boldsymbol{\ell}^{\text{def}}(M)|}{|\boldsymbol{\ell}^{\text{ref}}(M)|} \rightarrow 0 \quad \text{ as }  \quad M \rightarrow 0.
\end{align}
That is, even though the pattern has no periodicity, once these  sums become long enough, the distortion from $\boldsymbol{\ell}^{\text{ref}}$ to $\boldsymbol{\ell}^{\text{def}}$ is governed by the shape tensor $\mathbf{A}_{\text{eff}}(\xi)$. Essentially,  this result says that $\mathbf{A}_{\text{eff}}(\xi)$ quantifies the shape-change of the mechanism provided one averages over a large enough window.

We now make these ideas precise.  Consider any vector composed of a sum of connected panel edge vectors $\mathbf{s}_i$ on an aperiodic or random plane tiling 
\begin{equation}\label{eq:ellPT1}
\boldsymbol{\ell}^{\text{pt}}=\sum^{I}_{i = 1} \mathbf{s}_i.
\end{equation}
Note that there always exists a sum of connected panel centroid vectors with the following properties: 1) the head and/or tail   of the panel centroid vector belongs to a panel that contains the head and/or tail of a side length vector; 2) the sum of these vectors satisfies 
\begin{align}\label{eq:ellPT2}
    \boldsymbol{\ell}^{\text{pt}} = \sum^{I'}_{i'= 1} \mathbf{u}_{i'} + \boldsymbol{\delta} \mathbf{c}_1 +\boldsymbol{\delta} \mathbf{c}_2
\end{align}
where $\boldsymbol{\delta} \mathbf{c}_1$ starts at the tail of $\mathbf{s}_1$ and ends at the tail of $\mathbf{u}_1$, and where $\boldsymbol{\delta} \mathbf{c}_2$ starts at the head of $\mathbf{u}_{I'}$ and ends at the head of $\mathbf{s}_I$. An example of these vectors  overlaid onto the Penrose tiling is shown in Fig.\;\ref{fig:penrose}a. For future reference, we let $\mathcal{V}$ denote the collection of all vectors in the plane tiling formed by taking the head of the vector to be at a panel centroid and the tail at a  corner point of the same panel.  It follows trivially then that 
\begin{align}\label{eq:getLStar}
  |\boldsymbol{\delta} \mathbf{c}_1| \leq l^{\star} \quad \text{ and } \quad |\boldsymbol{\delta} \mathbf{c}_2| \leq l^{\star}  \quad \text{ where } \quad  l^{\star} = \max_{\mathbf{v} \in \mathcal{V}} |\mathbf{v}|.
\end{align}

After transforming the plane tiling to a parallelogram slit metamaterial using the separation tensor $\mathbf{D}$ (see Fig.\;\ref{fig:penrose}b) and then adding panels (see Fig.\;\ref{fig:penrose}c), we obtain the sum of connected edge vectors  
\begin{align}\label{eq:ellRefRandom}
\boldsymbol{\ell}^{\text{ref}} = \sum_{i = 1}^I \mathbf{s}_i + \sum_{i' = 1}^{I'} \boldsymbol{\delta} \mathbf{u}_{i'},
\end{align}
where each $\boldsymbol{\delta}\mathbf{u}_{i'}$ here satisfies 
\begin{align}\label{eq:geomIdent}
\boldsymbol{\delta}\mathbf{u}_{i'} = (\mathbf{D} - \mathbf{I}) \mathbf{u}_{i'} \quad \text{ for all } i' = 1,\ldots, I'
\end{align}
for the same elementary geometric reasons as stated in the main text.
The vector $\boldsymbol{\ell}^{\text{ref}}$ is illustrated on the Penrose metamaterial to the right in Fig.\;\ref{fig:penrose}c. Under a mechanism deformation (see Fig.\;\ref{fig:penrose}d),   each $\mathbf{s}_i$ deforms to $\mathbf{R}(-\xi) \mathbf{s}_i$ and each $\boldsymbol{\delta}\mathbf{u}_{i'}$ deforms to $ \mathbf{R}(\xi) \boldsymbol{\delta}\mathbf{u}_{i'}$. So $\boldsymbol{\ell}^{\text{ref}}$ deforms to $\boldsymbol{\ell}^{\text{def}}$ as 
\begin{align}\label{eq:ellDefRandom}
\boldsymbol{\ell}^{\text{def}} = \mathbf{R}(-\xi) \sum_{i=1}^{I} \mathbf{s}_i + \mathbf{R}(\xi) \sum_{i'=1}^{I'} \boldsymbol{\delta} \mathbf{u}_{i'}
\end{align}
This vector is illustrated to the right in the Penrose metamaterial in Fig.\;\ref{fig:penrose}d. To complete the description, observe by (\ref{eq:ellRefRandom}), (\ref{eq:geomIdent}) and (\ref{eq:ellDefRandom}) that 
\begin{align}
\boldsymbol{\ell}^{\text{ref}}=\sum_{i=1}^{I} \mathbf{s}_i+(\mathbf{D}-\mathbf{I})\sum_{i' = 1}^{I'} \mathbf{u}_{i'}, 
\qquad
\boldsymbol{\ell}^{\text{def}}=\mathbf{R}(-\xi)\sum_{i=1}^{I} \mathbf{s}_i+\mathbf{R}(\xi)(\mathbf{D}-\mathbf{I})\sum_{i' = 1}^{I'} \mathbf{u}_{i'}
\end{align}
Now apply (\ref{eq:ellPT1}) and (\ref{eq:ellPT2}) to these formula to get 
\begin{align}\label{eq:almostDoneRandom}
\boldsymbol{\ell}^{\text{ref}}= \mathbf{D}\boldsymbol{\ell}^{\text{pt}}+(\mathbf{I}-\mathbf{D})( \boldsymbol{\delta} \mathbf{c}_1 + \boldsymbol{\delta} \mathbf{c}_2), 
\qquad
\boldsymbol{\ell}^{\text{def}}=\big[\mathbf{R}(-\xi)  +\mathbf{R}(\xi)(\mathbf{D}-\mathbf{I})  \big] \boldsymbol{\ell}^{\text{pt}}   +\mathbf{R}(\xi)(\mathbf{I}-\mathbf{D})(\boldsymbol{\delta} \mathbf{c}_1 +  \boldsymbol{\delta} \mathbf{c}_2). 
\end{align}
Now bring in the definition of the shape tensor from the main text $\mathbf{A}_{\text{eff}}(\xi)=(\mathbf{R}(-\xi)-\mathbf{R}(\xi))\mathbf{D}^{-1}+\mathbf{R}(\xi)$ to obtain   \begin{align}\label{eq:almostDoneRandom1}
\boldsymbol{\ell}^{\text{def}} = \mathbf{A}_{\text{eff}}(\xi) \mathbf{D} \boldsymbol{\ell}^{\text{pt}} + (\mathbf{R}(-\xi) - \mathbf{A}_{\text{eff}}(\xi) \mathbf{D}) (\boldsymbol{\delta} \mathbf{c}_1 + \boldsymbol{\delta} \mathbf{c}_2) .
\end{align}

From here, we can show that $\boldsymbol{\ell}^{\text{def}} \approx \mathbf{A}_{\text{eff}}(\xi) \boldsymbol{\ell}^{\text{ref}}$.
Indeed, by (\ref{eq:ellDefRandom}) and (\ref{eq:almostDoneRandom1}) and standard estimates,   
\begin{equation}
    \begin{aligned}
        |\mathbf{A}_{\text{eff}}(\xi)  \boldsymbol{\ell}^{\text{ref}}-\boldsymbol{\ell}^{\text{def}}|&=|(\mathbf{A}_{\text{eff}}(\xi) - \mathbf{R}(-\xi))(\boldsymbol{\delta}\mathbf{c}_1+\boldsymbol{\delta}\mathbf{c}_2)  | \\
        &=|(\mathbf{R}(-\xi)-\mathbf{R}(\xi))(\mathbf{D}^{-1}-\mathbf{I})(\boldsymbol{\delta}\mathbf{c}_1+\boldsymbol{\delta}\mathbf{c}_2)|\\
        &=|2\sin{\xi}\mathbf{R}(\tfrac{\pi}{2})(\mathbf{D}^{-1}-\mathbf{I})(\boldsymbol{\delta}\mathbf{c}_1+\boldsymbol{\delta}\mathbf{c}_2)| \\
        &\leq 2|\mathbf{D}^{-1} - \mathbf{I}|(|\boldsymbol{\delta} \mathbf{c}_1| + |\boldsymbol{\delta}\mathbf{c}_2|) \\
        &\leq 4 \sqrt{2} l^{\star}
    \end{aligned}
\end{equation}
where $|\mathbf{D}^{-1} - \mathbf{I}| \leq \sqrt{2}$ since $\det \mathbf{D} >1$ and $\min_{\mathbf{e} \in \mathbb{S}^1} | \mathbf{D} \mathbf{e}| > 1$. Note, the last line uses (\ref{eq:getLStar}). It follows that 
\begin{equation}\label{eq:keyInequality}
    \begin{aligned}
        \frac{|\mathbf{A}_{\text{eff}}(\xi)\boldsymbol{\ell}^{\text{ref}}-\boldsymbol{\ell}^{\text{def}}|}{|\boldsymbol{\ell}^{\text{ref}}|}\leq\frac{4\sqrt{2}l^*}{|\boldsymbol{\ell}^{\text{ref}}|}
    \end{aligned}
\end{equation}
The key point is that  the numerator on the right in this result depends on the plane tiling used to construct the metamaterial  through $l^{\star}$ but not on the specific choices of $\boldsymbol{\ell}^{\text{pt}}$ and corresponding  $\boldsymbol{\ell}^{\text{ref}}$ and $\boldsymbol{\ell}^{\text{def}}$. Thus, if we choose any sufficiently long $\boldsymbol{\ell}^{\text{pt}}$ in this setup, the right side of this inequality becomes $\ll 1$. Physically, (\ref{eq:keyInequality}) furnishes  that $\boldsymbol{\ell}^{\text{def}}$ and $\mathbf{A}_{\text{eff}}(\xi) \boldsymbol{\ell}^{\text{ref}}$ only differ by the lengths of few panels, no matter how many edge vector are used in the construction. In other words, this estimate proves (\ref{eq:MellRefellDef}).

\section{Examples}

The general formula for the shape tensor of the designs in our  recipe is 
\begin{align}\label{eq:genShape}
    \mathbf{A}_{\text{eff}}(\xi) = ( \mathbf{R}(-\xi) - \mathbf{R}(\xi)) \mathbf{D}^{-1} + \mathbf{R}(\xi).
\end{align}
Kinematically, this tensor captures the effective  or cell averaged shape-change associated to the single DOF mechanism motion of a parallelogram slit design through  $\xi \in (\xi^{-}, \xi^+)$. As for design,  it encodes a four parameter catalogue of shape-changes obtained by specifying any  separation tensor $\mathbf{D} \in \mathbb{R}^{2\times2}$  that satisfies $\det \mathbf{D} > 1$ and $\min_{\mathbf{e} \in \mathbb{S}}| \mathbf{D} \mathbf{e}| > 1$.
The main text studies examples  within this design catalogue programmed to have axial shape-change through a diagonal shape tensor
\begin{align}\label{eq:diagShape}
\mathbf{A}_{\text{eff}}(\xi) = \sum_{i =1,2} \lambda_i(\xi) \mathbf{e}_i \otimes \mathbf{e}_i = \begin{pmatrix} \lambda_1(\xi) & 0 \\ 0 & \lambda_2(\xi) \end{pmatrix},   \end{align}
where $\mathbf{e}_1$ and $\mathbf{e}_2$ are the standard basis vectors on $\mathbb{R}^2$ in Cartesian coordinates. In what follows, we characterize all diagonal shape tensors furnished by our design recipe and illuminate their key properties.

\subsection{General parameterization of diagonal shape tensors}

\begin{figure}[b!]
    \centering
    \includegraphics[width=0.35\linewidth]{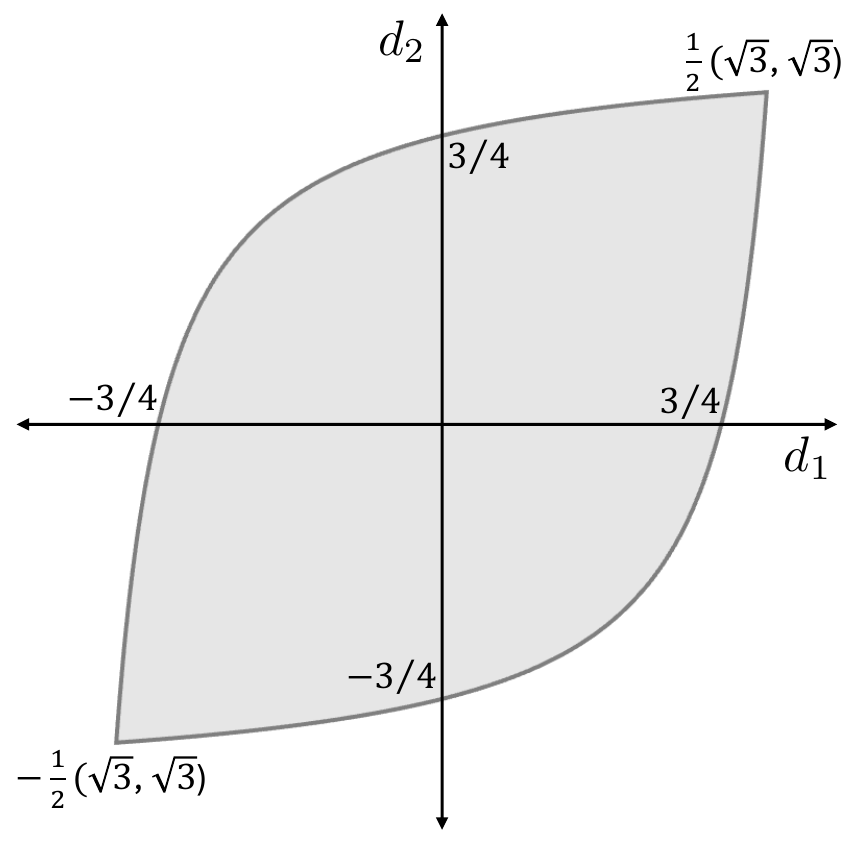}
    \caption{\textit{Region of validity for the parameters $d_1$ and $d_2$, characterizing the tensor $\mathbf{D}$ that leads to a metamaterial design with a diagonal shape tensor.}}
    \label{fig:diagTensorBounds}
\end{figure}

 We begin by deriving the general parameterization of $\mathbf{D}$ in (\ref{eq:genShape}) that leads to a diagonal shape tensor (\ref{eq:diagShape}). Observe  from (\ref{eq:Rxi}) that 
 \begin{equation}
 \begin{aligned}
&\mathbf{R}(\xi) = \cos \xi \begin{pmatrix}
    1 & 0 \\ 0 & 1
\end{pmatrix} + \sin \xi \begin{pmatrix} 
0 & -1  \\ 1 & 0
\end{pmatrix} = \cos \xi \mathbf{I} + \sin \xi \mathbf{R}(\tfrac{\pi}{2}), \\
&\mathbf{R}(-\xi) - \mathbf{R}(\xi) = 2 \sin \xi \begin{pmatrix} 0 &  1 \\ -1 & 0
\end{pmatrix} = 2 \sin \xi \mathbf{R}(-\tfrac{\pi}{2}).
 \end{aligned}
 \end{equation}
 In substituting these expressions into (\ref{eq:genShape}), 
  the shape tensor  can be written as 
\begin{align}
    \mathbf{A}_{\text{eff}}(\xi) = 2 \sin \xi \mathbf{R}(-\tfrac{\pi}{2}) \mathbf{D}^{-1} + \cos \xi \mathbf{I} + \sin \xi \mathbf{R}(\tfrac{\pi}{2}). 
\end{align}
From here, we see that $\mathbf{A}_{\text{eff}}(\xi)$ is a  diagonal shape tensor if and only if  $\mathbf{D}^{-1}$ cancels out the skew symmetric term $\sin \xi \mathbf{R}(\frac{\pi}{2})$. This is equivalent to prescribing 
\begin{align}
\mathbf{R}(-\tfrac{\pi}{2})\mathbf{D}^{-1} = \frac{1}{2}\mathbf{R}(-\tfrac{\pi}{2}) + \begin{pmatrix} d_1 & 0 \\ 0 & d_2 
\end{pmatrix}
\end{align}
for some $d_1, d_2 \in \mathbb{R}$, which becomes
\begin{align}\label{eq:diagD}
    \mathbf{D} = \frac{1}{1 + 4 d_1 d_2} \begin{pmatrix}  2 & 4 d_2 \\ -4 d_1 & 2\end{pmatrix}
\end{align}
after a bit of algebraic manipulation. We are now done, as substituting  (\ref{eq:diagD}) into (\ref{eq:genShape}) produces a diagonal shape tensor  (\ref{eq:diagShape}). The stretches in this expression are   of the form
\begin{equation}\label{eq:stretchesDiagonal}
\begin{aligned}
&\lambda_1(\xi) =\cos \xi + 2 d_1 \sin \xi  \quad \text{ and } \quad \lambda_2(\xi) = \cos \xi + 2 d_2 \sin \xi.
\end{aligned}
\end{equation}

\subsection{Region of validity}
In the main text, we introduced two constraints on the tensor $\mathbf{D}$, namely, $\det \mathbf{D} > 1$ and $\min_{\mathbf{e} \in \mathbb{S}^1} |\mathbf{D} \mathbf{e}| > 1$. These constraints ensure that this tensor separates all the panels in the plane tiling, so that the final metamaterial design has  no overlapping tiles. In the case of a diagonal shape tensor, for which $\mathbf{D}$ satisfies (\ref{eq:diagD}),  these conditions are equivalent to $d_1$ and $d_2$ such that
\begin{align}
    d_2 \in \Big( \frac{3-4d_1}{4(d_1-1)}, \frac{3 + 4d_1}{4(1 + d_1)}\Big), \quad d_1 \in \Big( -\frac{\sqrt{3}}{2}, \frac{\sqrt{3}}{2}\Big);
\end{align}
Fig.\,\ref{fig:diagTensorBounds} plots  this region of validity. 

\subsection{Poisson's Ratio} 

 An important property of a mechanism-based pattern is the geometric Poisson's ratio characterizing whether the shape change is auxetic or not. We now define this ratio for  any mechanism actuation $\xi$ of a diagonal shape tensor. 

A small perturbation $\delta \xi$ of a given actuation $\xi$ yields an effective displacement of the pattern $\mathbf{u}(\mathbf{y})$ satisfying
\begin{align}\label{eq:dispPoissons}
    \mathbf{u}(\mathbf{A}_{\text{eff}}(\xi)\mathbf{x}) = \mathbf{A}_{\text{eff}}(\xi + \delta \xi) \mathbf{x}  -  \mathbf{A}_{\text{eff}} (\xi)\mathbf{x}  \approx \delta \xi \mathbf{A}'_{\text{eff}}(\xi)  \mathbf{x}.
\end{align}
As is standard in continuum mechanics, the linear strain $\boldsymbol{\varepsilon}(\mathbf{y})$ is equal to the symmetric part of the displacement gradient   $\frac{1}{2} (\nabla \mathbf{u}(\mathbf{y}) + \nabla \mathbf{u}^T(\mathbf{y}))$. We conclude then from (\ref{eq:diagShape}) and (\ref{eq:dispPoissons}) that 
\begin{align}\label{eq:getStrain}
    \boldsymbol{\varepsilon}(\mathbf{A}_{\text{eff}}(\xi) \mathbf{x}) = \frac{\delta \xi}{2} \Big( \mathbf{A}_{\text{eff}}'(\xi)  \mathbf{A}_{\text{eff}}^{-1}(\xi) + \mathbf{A}_{\text{eff}}^{-T}(\xi)\big(\mathbf{A}_{\text{eff}}'(\xi)\big)^T\Big)  = \delta \xi \begin{pmatrix} \frac{\lambda_1'(\xi)}{\lambda_1(\xi)} & 0 \\ 0 & \frac{\lambda_2'(\xi)}{\lambda_2(\xi)} \end{pmatrix} .
\end{align}
The normal strains in this expression are  $\varepsilon_1 = \delta \xi \frac{\lambda_1'(\xi)}{\lambda_1(\xi)}$ and $\varepsilon_2 = \delta \xi \frac{\lambda_2'(\xi)}{\lambda_2(\xi)}$. Thus, by the conventional definition of the Poisson's ratio $\nu = -\frac{\varepsilon_2}{\varepsilon_1}$, we obtain  
\begin{align}\label{eq:poissonsDiag}
\nu(\xi) = -\frac{\lambda_2'(\xi) \lambda_1(\xi)}{\lambda_1'(\xi) \lambda_2(\xi)} = -\frac{(2 d_2 \cos \xi  - \sin \xi)(\cos \xi + 2 d_1 \sin \xi)}{(2 d_1 \cos \xi - \sin \xi)(\cos \xi + 2 d_2 \sin \xi)} .
\end{align}

The Poisson's ratio can also be defined  for general shape tensors $\mathbf{A}_{\text{eff}}(\xi)$ that are not diagonal. Simply use first formula for the strain $\boldsymbol{\varepsilon}(\mathbf{y})$ in (\ref{eq:getStrain}), then diagonalize this quantity to get the normal strains $\varepsilon_{\mathbf{v}}$ and $\varepsilon_{\mathbf{v}^{\perp}}$ associated to eigenvectors $\mathbf{v}$ and $\mathbf{v}^{\perp}$, and finally set $\nu_{\text{gen}} = - \frac{\varepsilon_{\mathbf{v}^{\perp}}}{\varepsilon_{\mathbf{v}}}$. We do not present calculations for this general setting here because it becomes cumbersome very quickly.

\vspace*{0.2cm}

\noindent \textbf{Conformal kirigami.} We can make the stretches in (\ref{eq:stretchesDiagonal}) equal by choosing $d_1 = d_2$. This parameterization yields a special class of auxetic kirigami that are purely dilation, called conformal kirigami.  All such patterns have a Poisson's ratio $\nu(\xi) = -1$ for all  $\xi$. In the main text, we showcase examples within this class for which $d_1 = d_2 = 0$.

\vspace*{0.2cm}

\noindent \textbf{Neutral Poisson's ratio kirigami.} Another interesting class of examples are patterns whose Poisson's ratio is zero at the reference state, corresponding to $\xi =0$. From (\ref{eq:poissonsDiag}), we have $\nu(0) = -\frac{d_2}{d_1}$ for a general diagonal shape tensor. So setting $d_2 = 0$ gives $\nu(0) = 0$ and  the approximations
\begin{align}
\nu(\xi)  \approx -\frac{1}{2d_1} \xi, \quad \lambda_1(\xi) \approx 1 + 2 d_1 \xi, \quad \lambda_2(\xi) \approx 1 - \frac{1}{2} \xi^2  
\end{align}
when $|\xi| \ll 1$. In these designs, stretching along the $\mathbf{e}_1$ direction ($2 d_2 \xi >0$) results in compression in the $\mathbf{e}_2$ direction, while compression in the $\mathbf{e}_1$ direction ($2 d_2 \xi <0$) also results in compression in $\mathbf{e}_2$. Thus, these patterns are auxetic when pushed and not auxetic when pulled.

\subsection{Generating quasi-random designs}
\begin{figure}
    \centering
    \includegraphics[width=0.7\linewidth]{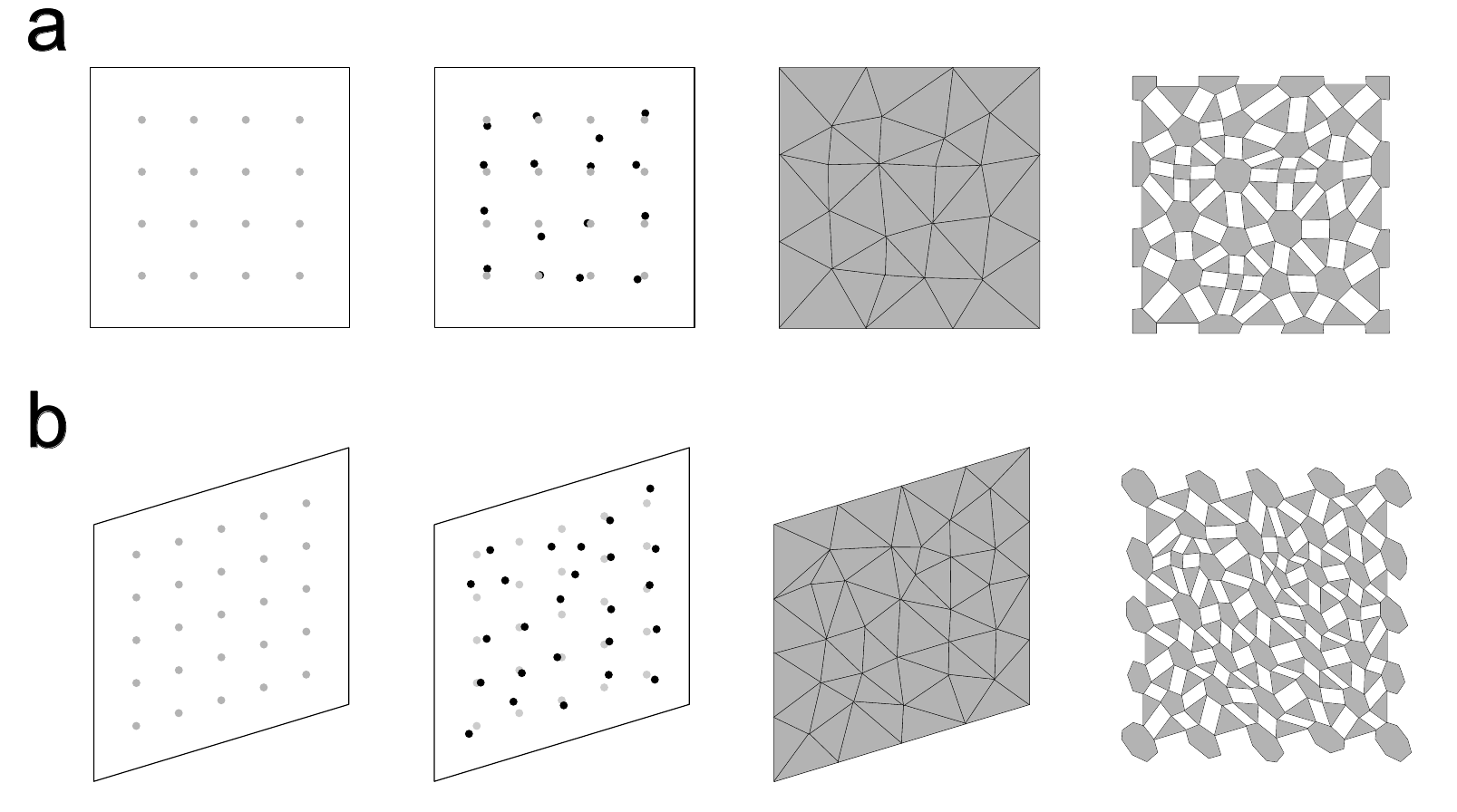}
    \caption{\textit{Construction of unstructured patterns. Examples for (a) conformal kirigami and (b) neutral Poisson’s ratio designs. From left to right: regular node arrangement within the domain; perturbed node positions obtained by applying random Gaussian perturbations; resulting panel geometry generated via a Delaunay triangulation; and the unstructured kirigami patterns obtained after applying the prescribed design tensor (Sample c and e in main text).}}
    \label{fig:DesignUnstructured}
\end{figure}

To demonstrate that the mechanism-based motion in our designs is agnostic to the underlying plane tiling, we generate unstructured patterns that share the same macroscopic shape-change as their periodic counterparts but possess entirely different microstructures. These patterns are generated by introducing random Gaussian perturbations to the node positions of an initially regular node arrangement within the domain. The perturbations are indicated by the black nodes in the center of Fig.\;\ref{fig:DesignUnstructured}a and b; they are generated from a distribution with zero mean and a standard deviation equal to 15\% of the characteristic spacing.  Next, the geometry is constructed by meshing the nodes into triangular panels using a Delaunay triangulation. In this way, the microstructure of the unstructured patterns before applying the  tensor $\mathbf{D}$ is markedly different from that of the structured samples.

By applying the design tensor corresponding to the \textit{conformal kirigami} examples ($d_1 = d_2 = 0$) to the meshed tiling shown in Fig.\;\ref{fig:DesignUnstructured}a, we obtain the sample shown in Fig.\;3c of the main text. Similarly, applying the design tensor of \textit{neutral Poisson's ratio} example ($d_1 = 0.15,\; d_2 = 0$) to the meshed gray tiling in Fig.~\ref{fig:DesignUnstructured}b yields the sample shown in Fig.\;3e. Boundary panels are generated by treating the meshed domain as a unit cell of an underlying periodic tiling; neighboring panels are added as if the design were periodically extended. This construction allows the counter-rotating mechanism to be applied consistently at the boundaries.

In both cases, the random seeds had to be modified slightly to ensure  that the resulting pattern is physically admissible, i.e., that  panels do not overlap  and that all slits remain convex in the reference configuration.  This involved the slight manipulation of a handful of nodes  in each case, which of course is not strictly a random process. Hence, we refer to these designs as quasi-random patterns in the main text.

\section{Fabrication,  experimental procedures and validation}

In this section, we provide additional details on the design and fabrication of the physical specimens, and on the experiments conducted in this work.


\subsection{Fabrication}

In the literature, planar kirigami patterns are typically monolithic, with panels connected by small elastic hinges produced via 3D printing or laser cutting. However, due to the zero-mode deformation of the structures studied here, we fabricate specimens using acrylic panels and push-in rivets acting as pin joints. The design and fabrication steps are reported below.
\begin{figure}[htb]
    \centering
    \includegraphics[width=0.99\linewidth]{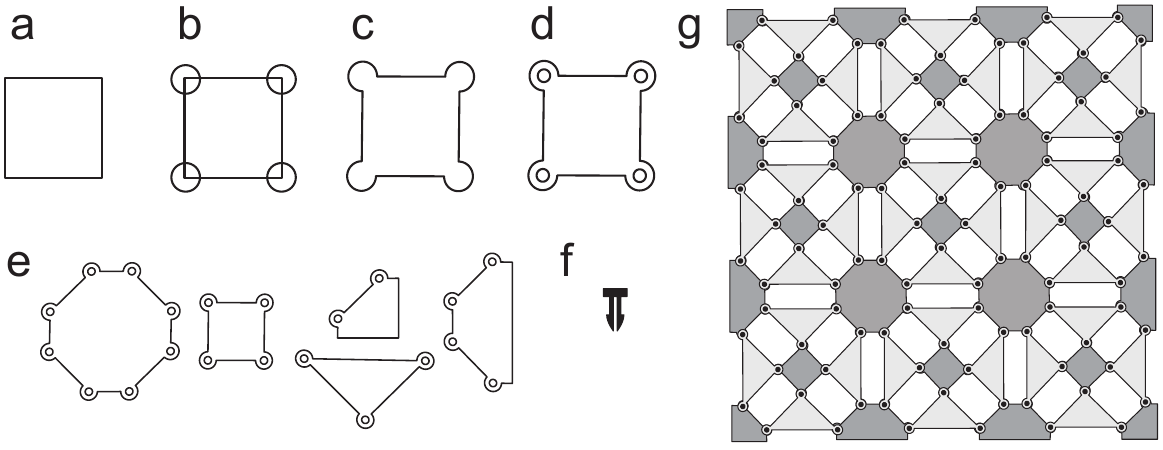}
    \caption{\textit{Design workflow for all specimens. (a) Outline of a theoretically designed panel. (b--c) Intermediate steps for the MATLAB-generated outline of (d) the pin-jointed version of the same panel. Illustration of (e) the unique panels required for the conformal kirigami pattern, and (f) one of the push-in rivets. (g) Assembled $3 \times 3$ structure via rivets; note that panels are arranged in only two layers: dark gray at the bottom and light gray at the top.}}
    \label{fig:Fig_SM_A1}
\end{figure}

The fabrication process is identical for all geometries, and illustrated for one of our patterns in Fig.~\ref{fig:Fig_SM_A1}.
\begin{itemize}
    \item We import the vertex coordinates for each panel of a unit cell into MATLAB (Fig.~\ref{fig:Fig_SM_A1}a) and convert each panel into a MATLAB \texttt{polyshape} object.
    \item We center a 7~mm diameter circle at each joint (Fig.~\ref{fig:Fig_SM_A1}b) and apply the \texttt{union} function to obtain an expanded shape (Fig.~\ref{fig:Fig_SM_A1}c).
    \item We place a 3~mm diameter circle at each joint (Fig.~\ref{fig:Fig_SM_A1}d) and use the \texttt{subtract} function on the previously obtained shape (Fig.~\ref{fig:Fig_SM_A1}c) to create the holes for the push-in rivets.
    \item We prepare all unique panels required for the assembly, as shown in Fig.~\ref{fig:Fig_SM_A1}e.
    \item These drawings are used to laser-cut individual panels from 1.5~mm-thick acrylic (McMaster-Carr part number: 8650K11). We operate the 80W laser cutter (Epilog FusionPro 30) at 15\% power, 100\% speed, and 100\% frequency.
    \item Once the panels are cut, we assemble them using push-in rivets (Fig.~\ref{fig:Fig_SM_A1}f, McMaster-Carr part number: 90136A630). A key aspect of this assembly is ensuring that cells are arranged in only two layers, as distinguished by the dark and light gray shading in Fig.~\ref{fig:Fig_SM_A1}g.
\end{itemize}

\subsection{Experimental procedures}

We now provide details regarding the kinematic experiments. As the objective is to measure the stretch ratio of the geometries, we take distance measurements between selected points of the specimens using a ruler. Before we conduct the experiments, it is necessary to establish a coordinate system to track these distances as the geometry deforms. We establish this system by selecting reference lines passing through specific pin joints on each geometry, such that these lines remain parallel to either the horizontal or vertical direction during deformation. 

\begin{figure}[htb]
    \centering
    \includegraphics[width=0.99\linewidth]{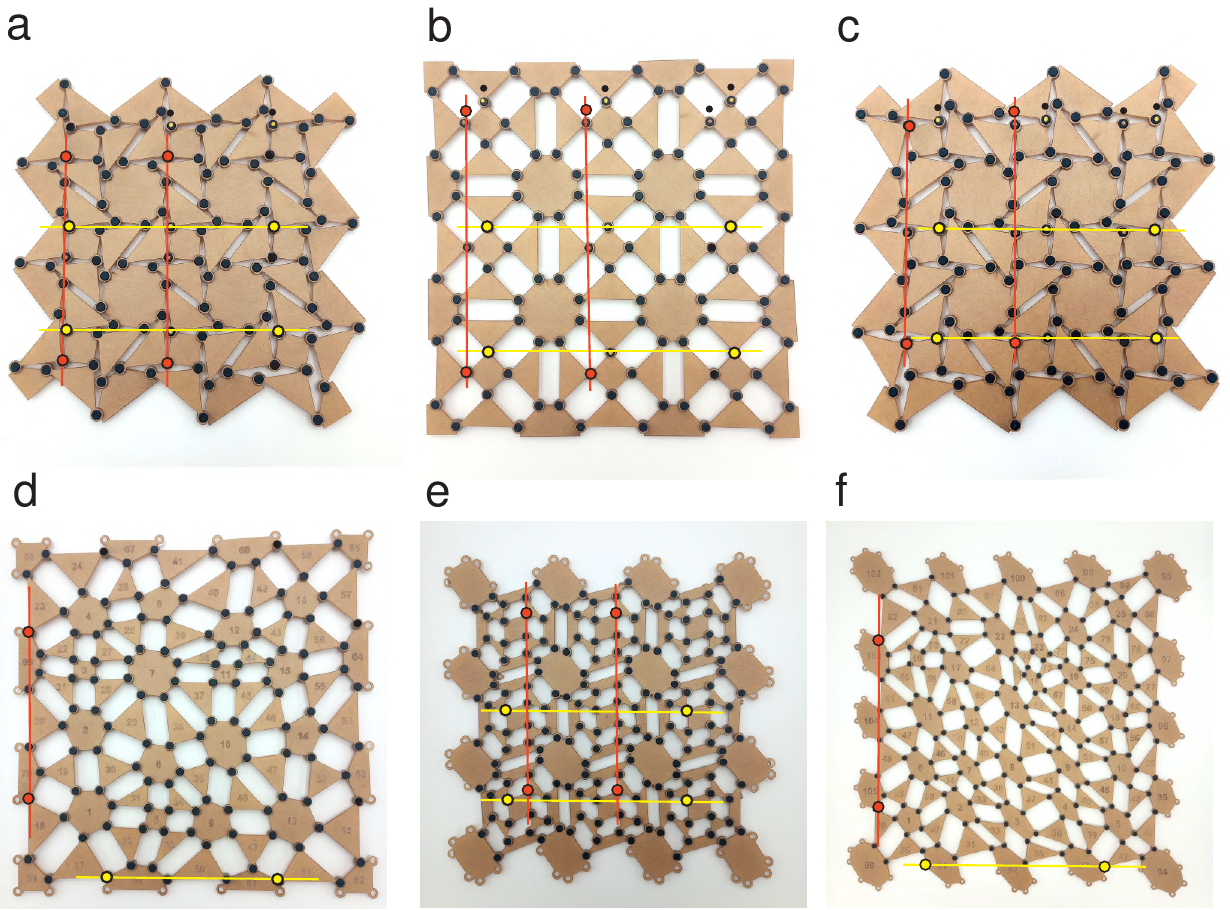}
    \caption{\textit{Reference lines for establishing the coordinate system in various samples. The structured conformal kirigami specimen in its (a) first collapsed state, (b) fully expanded state, and (c) second collapsed state. (d) The unstructured conformal specimen in its fully expanded state. Neutral Poisson’s ratio kirigami in the fully expanded state for both (e) structured and (f) unstructured geometries.}}
    \label{fig:Fig_SM_A1_Axes}
\end{figure}

To illustrate this process, we use the conformal kirigami specimen shown in Fig.~\ref{fig:Fig_SM_A1_Axes}a and \ref{fig:Fig_SM_A1_Axes}c in its two collapsed states, and in Fig.~\ref{fig:Fig_SM_A1_Axes}b in its fully expanded state. The two yellow horizontal lines --- one at the center of the specimen and another near the boundary --- remain parallel across all states. Similarly, the two orange lines remain parallel throughout the deformation. Once the coordinate system is established, these lines are used to determine horizontal and vertical stretch by measuring and averaging the distances between points along each line. For the remaining geometries, the fully expanded states and the points used to establish coordinates are shown in Fig.~\ref{fig:Fig_SM_A1_Axes}d--f. For the unstructured samples (Fig.~\ref{fig:Fig_SM_A1_Axes}d, f), only one set of axes is used near the boundary, as these geometries lack points near the center that remain parallel across different states.

We begin measurements from one collapsed state and take readings by rotating the panels by arbitrary increments until reaching the second collapsed state. The entire process is repeated three times. Reference lengths are determined from measurements taken when the structure is in its fully expanded state. 

\begin{figure}[h!]
    \centering
    \includegraphics[width=0.99\linewidth]{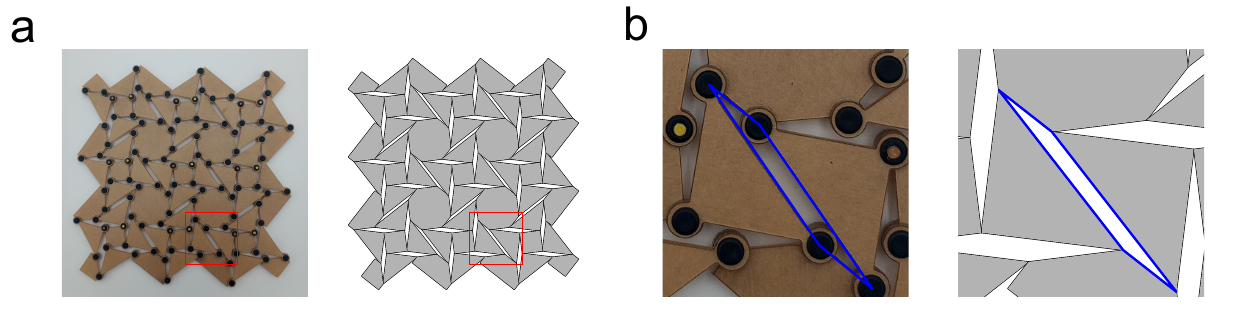}
    \caption{\textit{Validation of the pin-induced geometric obstruction that limits the experimental actuation range. (a) Left: experimental limit state with the most-closed slit highlighted. Right: corresponding theoretical configuration with actuation $\xi=\xi_{\text{exp}}^-$. (b) Detailed comparison of the highlighted slit in experiment (left) and theory (right).}}
    \label{fig:Validation}
\end{figure}

\subsection{Validation}

It is important to note that, since our panels feature hinges of finite size, fully collapsing the physical specimens is impossible. For this reason, our measurements cannot capture the tail ends of the theoretical lines in Fig.~3 of the main manuscript. 

 Specifically, in the experiments, neighboring panels are connected by circular pins of finite radius, which allow free in-plane rotation but occupy a nonzero area. As the structure approaches its theoretical limit states, further rotation would require the slits to fully close. However, prior to complete closure, the pins come into contact with adjacent panel edges, as shown in Fig.~\ref{fig:Validation}a. This contact prevents additional rotation and limits the experimentally accessible deformation to $\xi=\xi_{\text{exp}}^-$, with $|\xi_{\text{exp}}^-|<|\xi^-|$. In contrast, the theoretical design assumes ideal point hinges and therefore permits full slit closure. However, when we  actuate the theoretical sample to $\xi = \xi^{-}_{\text{exp}},$ as shown to the right in Fig.\;\ref{fig:Validation}a-b,  the  experiment and theory are in near perfect agreement. Thus, the discrepancy in main text is  entirely attributable to this pin-induced geometric obstruction, and not to the underlying kinematics.

\section{Compact kirigami}

The end of the main text briefly discussed potential generalizations to our design recipe, the first of which is \textit{compact kirigami}. The basic idea for these designs is that, if we can separate the plane tiling in such a way that all the slits are rectangles, then the mechanism actuation will yield two compact and fully closed states as $\xi \rightarrow \xi^{\pm}$. To see this, recall that (\ref{eq:xiPM}) identifies the extents $\xi^{\pm}$ of the mechanism actuation. In the special case that  the slits are rectangles, the slit angles $\theta_i = \frac{\pi}{2}$ for all $i$ and so   $\xi^{+} = \frac{\pi}{4}$ and $\xi^{-} = - \frac{\pi}{4}$. As the actuation evolves these  angles  with $\xi$ via  $\frac{\pi}{2}  \rightarrow \frac{\pi}{2} + 2\xi$ (recall $\theta_0 \rightarrow \theta_0 +2\xi$ in Fig.\;\ref{fig:SingleSlit}),  all the slits close simultaneously when $\xi = \xi^{\pm} = \pm \frac{\pi}{4}$.
\begin{figure}[h!]
    \centering
    \includegraphics[width=0.8\linewidth]{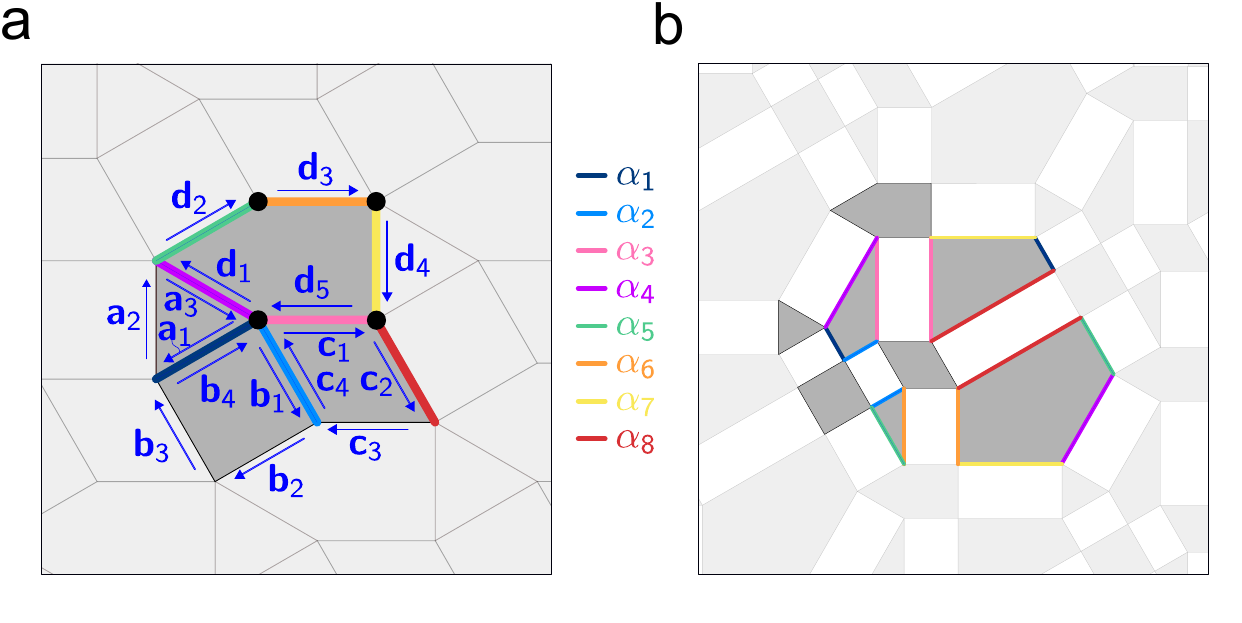}
    \caption{\textit{Designing compact kirigami with rectangular slits. (a) Initial plane tiling --- the emphasized panels form the unit cell of the plane tiling; the four black dots indicate the added panels in the unit cell after transformation; the eight colored edges, labeled by $\alpha_i$, $i =1,\ldots 8$, indicate the slits in the unit cell after transformation; the vectors $\mathbf{a}_1,\ldots, \mathbf{d}_5$ label these edges. (b) Solving the compatibility conditions outlined below leads to a pattern with rectangular slits.} }
    \label{fig:DesignCompactKiri}
\end{figure}

We now outline a strategy to design compact kirigami patterns with rectangular slits, starting from any plane tiling.  Fig.\;\ref{fig:DesignCompactKiri}a shows one such plane tiling. When we separate all the panels and add new panels according to our recipe, the nodes in this tiling become the added panels, while the edges become parallelogram slits. Note that the four black dots, four shaded panels, and eight colored edges form a unit cell of eight panels and eight slits after the transformation. We enforce that the slits are rectangles as follows: notice that  the first slit, labeled by $\alpha_1$, must have its four sides  be identified by the vectors $ \mathbf{a}_1, \alpha_1 \mathbf{R}(\frac{\pi}{2}) \mathbf{a}_1, \mathbf{b}_4,$ and $\alpha_1 \mathbf{R}(\frac{\pi}{2}) \mathbf{b}_4$ for some $\alpha_1 >0$. All the other slits posses analogous geometric relationships based on the labeling scheme in the figure. Of course, this prescription cannot be   arbitrary --- it must solve the  four loop compatibility conditions associated to generating the  new panels, namely, 
\begin{equation}\label{eq:compatCompact}
\begin{aligned}
   &\mathbf{R}(\tfrac{\pi}{2})\big(  \alpha_1 \mathbf{b}_4+ \alpha_4 \mathbf{a}_3 + \alpha_3 \mathbf{d}_5 +  \alpha_2 \mathbf{c}_4\big)  = \mathbf{0}, \\
   &\mathbf{R}(\tfrac{\pi}{2}) \big( \alpha_5 \mathbf{d}_2 + \alpha_6 \mathbf{c}_3 + \alpha_2 \mathbf{b}_1) = \mathbf{0}, \\
   & \mathbf{R}(\tfrac{\pi}{2}) \big( \alpha_6 \mathbf{d}_3 + \alpha_7 \mathbf{a}_2 + \alpha_5 \mathbf{b}_2 + \alpha_8 \mathbf{c}_2+\alpha_4 \mathbf{d}_1\big) = \mathbf{0}, \\
   &\mathbf{R}(\tfrac{\pi}{2}) \big( \alpha_7 \mathbf{d}_4 + \alpha_3 \mathbf{c}_1 + \alpha_8 \mathbf{b}_3+ \alpha_1 \mathbf{a}_1 \big) = \mathbf{0}. 
\end{aligned}
\end{equation}
from some $\alpha_i > 0$, $i =1,\ldots, 8$. In fact, one can readily check that these constraints are also sufficient for generating a rectangular-slit pattern. Indeed, after solving for (\ref{eq:compatCompact}), we only need to verify that the newly generated panels are convex polygons, which involves enforcing a consistent sign when taking the  cross-product of consecutive edge vectors. These conditions turn out to trivially hold when each  $\alpha_i >0$.

 The compatibility conditions in (\ref{eq:compatCompact})  describe a  linear system of equations in the vector $\boldsymbol{\alpha}$, listing the  $\alpha_i$ components.   Succinctly,  we  obtain a rectangular-slit pattern if and only if we find a vector $\boldsymbol{\alpha}$ such  that  
\begin{align}
\mathbf{C} \boldsymbol{\alpha} = \mathbf{0} \quad \text{ and } \quad \boldsymbol{\alpha} > 0,
\end{align}
where  $\mathbf{C}$ depends only on the edge vectors of the initial plane tiling, i.e.,  $\mathbf{a}_1, \ldots, \mathbf{d}_5$ in the case in Fig.\;\ref{fig:DesignCompactKiri}a.
Although we arrived at this statement using  the specific example in  the figure,   it clearly generalizes to any initial plane tiling --- the dimension and  entries   of the matrix $\mathbf{C}$ will change, but we will still need to find a positive null-space vector to obtain a solution.  An obvious question, then, is whether these  equations always admit solutions. On this point, observe for the plane tiling in Fig.\;\ref{fig:DesignCompactKiri}a that   
\begin{equation}\label{eq:edgeSymmetries}
\begin{aligned}
&\mathbf{b}_4 + \mathbf{a}_1 = \mathbf{0}, && \mathbf{b}_1 + \mathbf{c}_4 = \mathbf{0}, && \mathbf{c}_1 + \mathbf{d}_5 = \mathbf{0}, && \mathbf{d}_1 + \mathbf{a}_3 = \mathbf{0},\\
&\mathbf{d}_2 + \mathbf{b}_2  =\mathbf{0}, && \mathbf{d}_3 + \mathbf{c}_3  =\mathbf{0}, &&\mathbf{a}_2 + \mathbf{d}_4 = \mathbf{0}, && \mathbf{b}_3 + \mathbf{c}_2 = \mathbf{0},  
\end{aligned}
\end{equation}
since each edge is associated to two edge vectors of opposite direction. As a result, two of the eight equations in  (\ref{eq:compatCompact}) are degenerate --- simply  add  them all up and use (\ref{eq:edgeSymmetries}) --- and thus the null-space of $\mathbf{C}$ for this example is at least two dimensional. It's clear that statement also  holds  for general plane tilings, namely, that their  corresponding  matrix $\mathbf{C}$ has at least a two dimensional null-space.  In particular,  we will always have two vectors labeling each edge that cancel out when adding all of the compatibility equations together. Unfortunately, it seems harder to gauge   whether such null-spaces possess a vector $\boldsymbol{\alpha} > 0$ in general;  we leave this general question open for now.

To end this section, we have  verified by a direct computation of $\mathbf{C}$ that the plane tiling in Fig.\;\ref{fig:DesignCompactKiri}a does possess  positive vectors in its null-space.  One such example is highlighted in Fig.\;\ref{fig:DesignCompactKiri}b; this example also serves as our  illustration of compact kirigami in the main text. 

\section{Non-planar kirigami}\label{sec:infiniteAnalysis}

In this section, we verify that panels can be   added to an initial 2D kirigami design to produce a pattern with a single DOF bulk mechanism in 3D. First, we show by way of a counting argument that the original designs are too floppy when allowed to deform out-of-plane. Then,  we perform a kinematic analysis on a specific added-panel  design that shows it has a single DOF mechanism, with no other spurious floppy modes of deformation.   While our approach in part focuses on a specific example, the underlying strategy  can be applied any parallelogram slit design. 

\subsection{Counting arguments for floppy-to-rigid designs}

When a parallelogram slit kirigami pattern is permitted to deform out-of-plane, it becomes much floppier. As useful counting argument to this effect, one can supply a bound for the number of mechanisms in planar kirigami or non-planar kirigami and origami by simply comparing the  number of panel rotations (DOFs) in these patterns to the number of corresponding compatibility conditions (constraints). In 2D, the bound is 
\begin{align}\label{eq:2DCount}
    N_{\text{mech.}} \geq  N_{\text{panels}} - 2 N_{\text{slits}}.
\end{align}
Each slit introduces two constraints (recall (\ref{eq:loopCompat})), while each 2D panel rotation is   1 DOF. 
In 3D, we instead have 
\begin{align}\label{eq:3DCount}
  N_{\text{mech.}} \geq   3N_{\text{panels}} - 3 N_{\text{slits}} - 2 N_{\text{creases}},
\end{align}
as  3D panel rotations are 3 DOFs and  the corresponding loop conditions are 3D equations, and so  3 constraints.   The creases also introduce constraints in this formula ---  adjacent panels are compatible along a crease if and only if  $\mathbf{R} \mathbf{s} = \mathbf{Q} \mathbf{s}$ where $\mathbf{s}$ is a vector tangent to the   shared edge and $\mathbf{Q}, \mathbf{R}$ are the two panels rotations. Equations of this type are only two constraint, as indicated in (\ref{eq:3DCount}), because $|\mathbf{R} \mathbf{s}| = |\mathbf{Q}\mathbf{s}| = |\mathbf{s}|$ holds  trivially.

\begin{figure}[h!]
    \centering
    \includegraphics[width=0.99\linewidth]{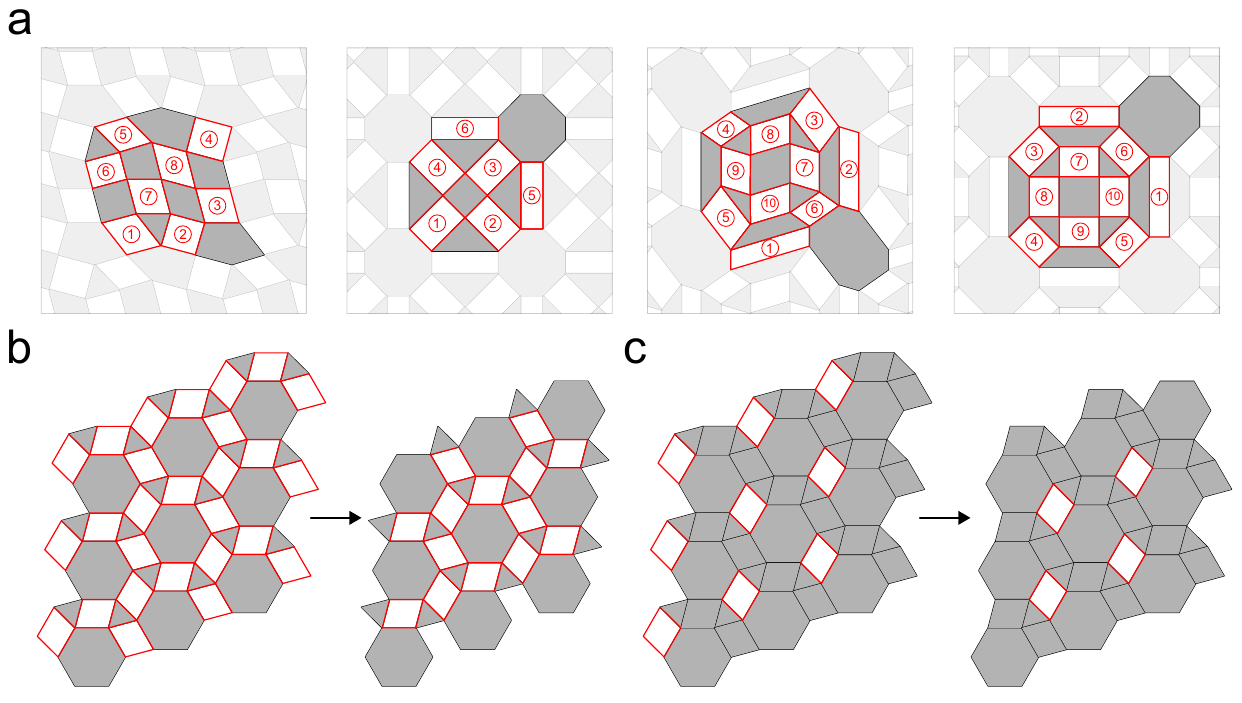}
    \caption{\textit{Counting constraints and DOFs in parallelogram-slit krigami. (a) Each unit cell produced by our design recipe contains an equal numbers of slits and panels. (b) A $3\times3$ array with a unit cell of one hexagon and two trianglar panels; (c) the same array with two slits per cell converted to panels.  Pruning the boundary, as shown, always eliminates more constraints than DOFs.  } }
    \label{fig:CountNonPlanarKiri}
\end{figure}

In both (\ref{eq:2DCount}) and (\ref{eq:3DCount}), this counting argument is a lowerbound   on the number of mechanisms because the constraints listed in the count can be (and often are) redundant. This is the case for all parallelogram slit designs in 2D.  Fig.\;\ref{fig:CountNonPlanarKiri}a shows a variety of such designs. In every case, the unit cell has the same number of slits as panels, and thus  more constraints per cell in 2D than DOFs. However, each possess a single DOF 2D bulk mechanism even though this count suggests otherwise.

The situation is vastly different in 3D: the  parallelogram slit designs are now isostatic  --- at a cell level, they  have the same number of constraints as DOFs (compare (\ref{eq:2DCount}) to  (\ref{eq:3DCount})).  Much like 2D Maxwell lattices \cite{lubensky2015phonons}, this balance means that any pattern with a  finite number of unit cells is guaranteed to be floppy; in fact, the lower bound on the the number of mechanisms grows linearly with $N$ for an $N \times N$ pattern. Fig.\;\ref{fig:CountNonPlanarKiri}b illustrates this basic point. The left pattern shows the unit cell repeated as a $3\times3$ array.   The count for this pattern is  $3N_{\text{panel}} - 3N_{\text{slits}} = 0$ only because we are including slits and panels on the boundary that are  unconstrained by neighboring unit cells. When we prune the boundary, as shown to the right, the count becomes $3N_{\text{panels}} - 3N_{\text{slits}} = 33$. Thus, this pattern is extremely floppy in 3D.

Replacing some of the  slits  with panels fundamentally alters this calculation. Fig.\;\ref{fig:CountNonPlanarKiri}c shows the same pattern as Fig.\;\ref{fig:CountNonPlanarKiri}b, but with two  slits per cell replaced by panels. As each added panel produces four creases, the count per cell in this setting is  $(  3N_{\text{panels}} - 3 N_{\text{slits}} - 2 N_{\text{creases}})/\text{cell} = -2N_{\text{added panel}}$, which is $-4$ for the pattern shown. So, even though finite size samples have extra DOFs due to their boundary, it is not enough to overcome this per-cell imbalance. In particular, the pruned $3\times3$ array on the right in Fig.\;\ref{fig:CountNonPlanarKiri}c satisfies $3N_{\text{panels}} - 3 N_{\text{slits}} - 2 N_{\text{creases}} = 3(37) -3(4)- 2(52) = -5$, suggesting that the pattern has been rigidified.

\subsection{Verifying a mechanism-based response}

\begin{figure}[h!]
    \centering
    \includegraphics[width=0.9\linewidth]{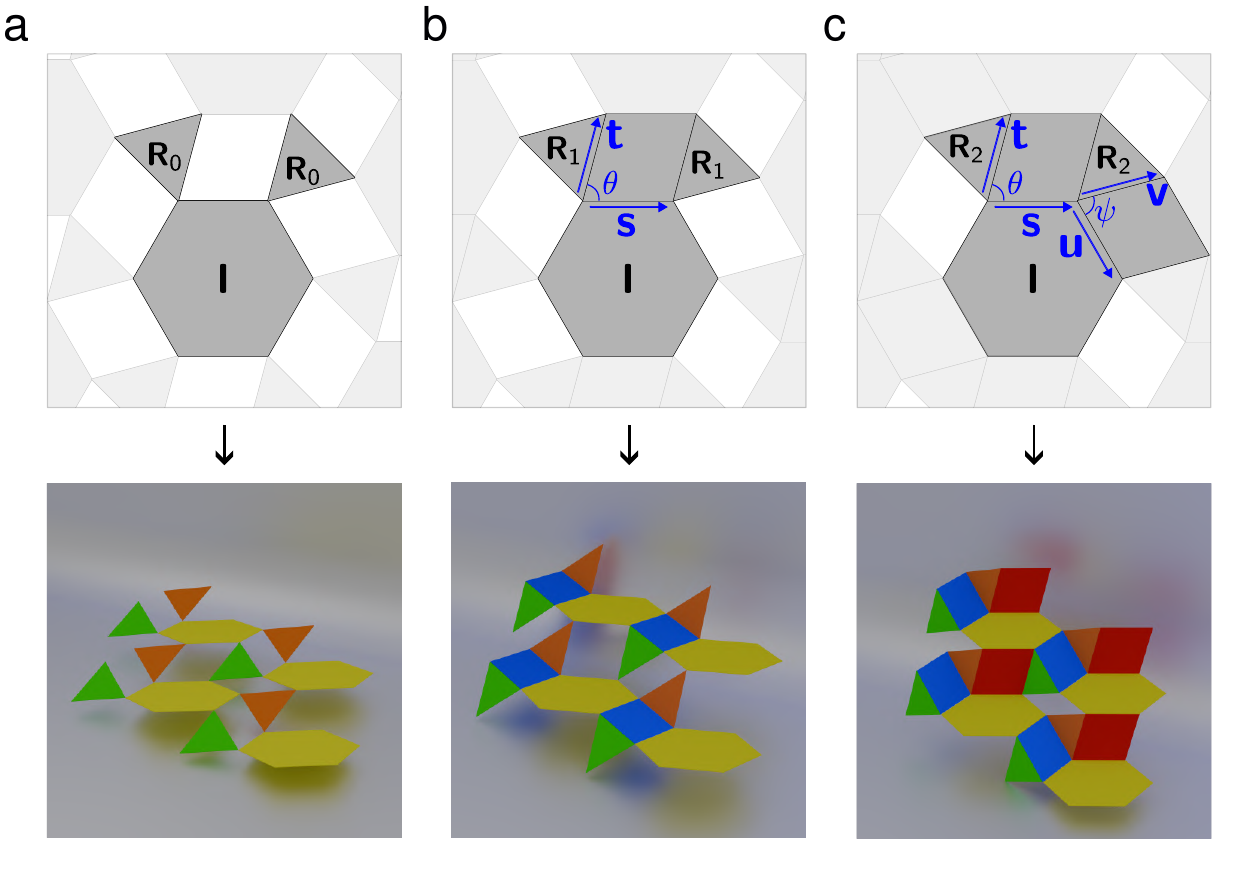}
    \caption{\textit{Bulk mechanisms of non-planar kirigami. (a) A pattern generated from a plane tiling of hexagons --- its bulk mechanism is obtained by rotating each triangle generically by $\mathbf{R}_0$ while keeping the hexagons fixed. (b) Replacing one slit in the unit cell with a panel constrains the kinematics of the bulk mechanism --- the rotation $\mathbf{R}_1$ must preserve the sector angle $\theta$. (c) Replacing another slit constrains the bulk mechanism  kinematics further --- the rotation $\mathbf{R}_2$ needs to preserve two sector angles, $\theta$ and $\psi$.}}
    \label{fig:DesignNonPlanarKiri}
\end{figure}

\vspace*{0.2cm}
\noindent  \textbf{Bulk mechanisms.}\;While the previous counting argument suggests a floppy-to-rigid transition, related to the added panels, we now  show that the 2D kirigami patterns and their added panel variants always posses 3D bulk mechanism, where the panels deform out-of-plane with rotations that repeat from cell to cell. 

Fig.\;\ref{fig:DesignNonPlanarKiri}a shows  a parallelogram slit design obtained via  our design recipe, starting from  a plane tiling of regular hexagons. In 2D, this pattern's mechanism deformations are given by rotating each hexagon by the 2D rotation   $\mathbf{R}(\xi)$ and counter-rotating the triangles by $\mathbf{R}(-\xi)$ for an angle $\xi \in (\xi^{-}, \xi^+)$. In 3D, we can replace  $\mathbf{R}(\xi)$ and $\mathbf{R}(-\xi)$ by the 3D rotations $\mathbf{I}$ and $\mathbf{R}_0 \in SO(3)$, as shown; the panels remain compatible with the parallelogram slits since the local loop conditions, i.e., (\ref{eq:loopCompat}), do not fundamentally change in this extension to 3D.  The main differences  in going from 2D to 3D are twofold: 1) this bulk mechanism is now described by 3 DOFs  since $\mathbf{R}_0$ is a general rotation; 2)  the pattern now has many additional mechanisms, per the  the prior counting argument.

Adding panels rigidifies that pattern. However, they retain a bulk mechanism but with a more restricted kinematics. Indeed, Fig.\;\ref{fig:DesignNonPlanarKiri}b highlights the case of replacing one  slit per cell. In this setting, applying the rotations $\mathbf{I}$ and $\mathbf{R}_1 \in SO(3)$ to the the panels, as shown, preserves the loop compatibility of every slit. However, the added panel introduces a non-trivial constraint ---  the sector angle $\theta$ in the figure must remain constant under this actuation. To characterize this constraint, it is useful to introduce the general parameterization of a rotation in $SO(3)$ in terms of a non-zero vector $\mathbf{a} \in \mathbb{R}^3$ and angle $\alpha$ 
\begin{align}
   \mathbf{R}_{\mathbf{a}}(\alpha) = \hat{\mathbf{a}} \otimes \hat{\mathbf{a}} + \cos \alpha \big( \mathbf{I} -  \hat{\mathbf{a}} \otimes \hat{\mathbf{a}} \big) + \sin \alpha (\hat{\mathbf{a}} \times)   
\end{align}
where $\hat{\mathbf{a}} = \tfrac{\mathbf{a}}{|\mathbf{a}|}$ and $(\mathbf{a} \times)$ is a tensor that satisfies $(\mathbf{a} \times)\mathbf{b} = \mathbf{a} \times \mathbf{b}$ for all $\mathbf{b} \in \mathbb{R}^3$. From here, note that $\theta$ is constant under the rotation $\mathbf{R}_1$ if and only if 
\begin{align}\label{eq:perserveAngle}
    \mathbf{s} \cdot \mathbf{t} = \mathbf{s} \cdot \mathbf{R}_1 \mathbf{t},  
\end{align}
which is equivalent to  
\begin{align}
    \mathbf{R}_1 = \mathbf{R}_{\mathbf{s}}(\omega) \mathbf{R}_{\mathbf{t}}(\gamma) 
\end{align}
for some angles $\omega$ and $\gamma$.
 So, the rotation $\mathbf{R}_1$ is not generic and  instead has two DOFs parameterized by $\omega$ and $\gamma$.

We can further rigidify the pattern by replacing a second slit in each cell with a panel. As shown in  Fig.\;\ref{fig:DesignNonPlanarKiri}c, the panel rotations $\mathbf{I}$ and $\mathbf{R}_2 \in SO(3)$  now need to preserve two sector angles, $\theta$ and $\psi$. Similar to (\ref{eq:perserveAngle}),  this means that $\mathbf{R}_2$  must satisfy 
\begin{align}
    \mathbf{s} \cdot \mathbf{t} = \mathbf{s} \cdot \mathbf{R}_2 \mathbf{t} \quad \text{ and } \quad \mathbf{u} \cdot \mathbf{v} = \mathbf{u} \cdot \mathbf{R}_2 \mathbf{v} ; 
\end{align}
equivalently, 
\begin{align}\label{eq:R2Param}
    \mathbf{R}_2 = \mathbf{R}_{\mathbf{s}}(\omega) \mathbf{R}_{\mathbf{t}}(\gamma) = \mathbf{R}_{\mathbf{u}}(\eta) \mathbf{R}_{\mathbf{v}}(\zeta) 
\end{align}
for angles $\omega, \gamma, \eta$ and $\zeta$. The latter constraint is exactly the same as the local compatibility condition for folding origami at a four-fold vertex. It is well known that  these types of equations  generically have two families of solutions,  
constraining three of the four angles via $(\omega, \gamma, \eta, \zeta) = (\omega, \gamma^{\pm}(\omega), \eta^{\pm}(\omega), \zeta^{\pm}(\omega))$ where   $\omega$ parameterizes the mechanism and the $\pm$ indicate the two distinct mountain-valley assignments of this bulk mechanism. While such parameterizations can be made explicit, they are cumbersome to write out; we refer the interested reader to \cite{feng2020designs,foschi2022explicit} where these kinematics are well-developed. All told, the bulk mechanism reduces to a single DOF when we replace a second slit in the unit cell with a panel. 
\begin{figure}
    \centering
    \includegraphics[width=0.9\linewidth]{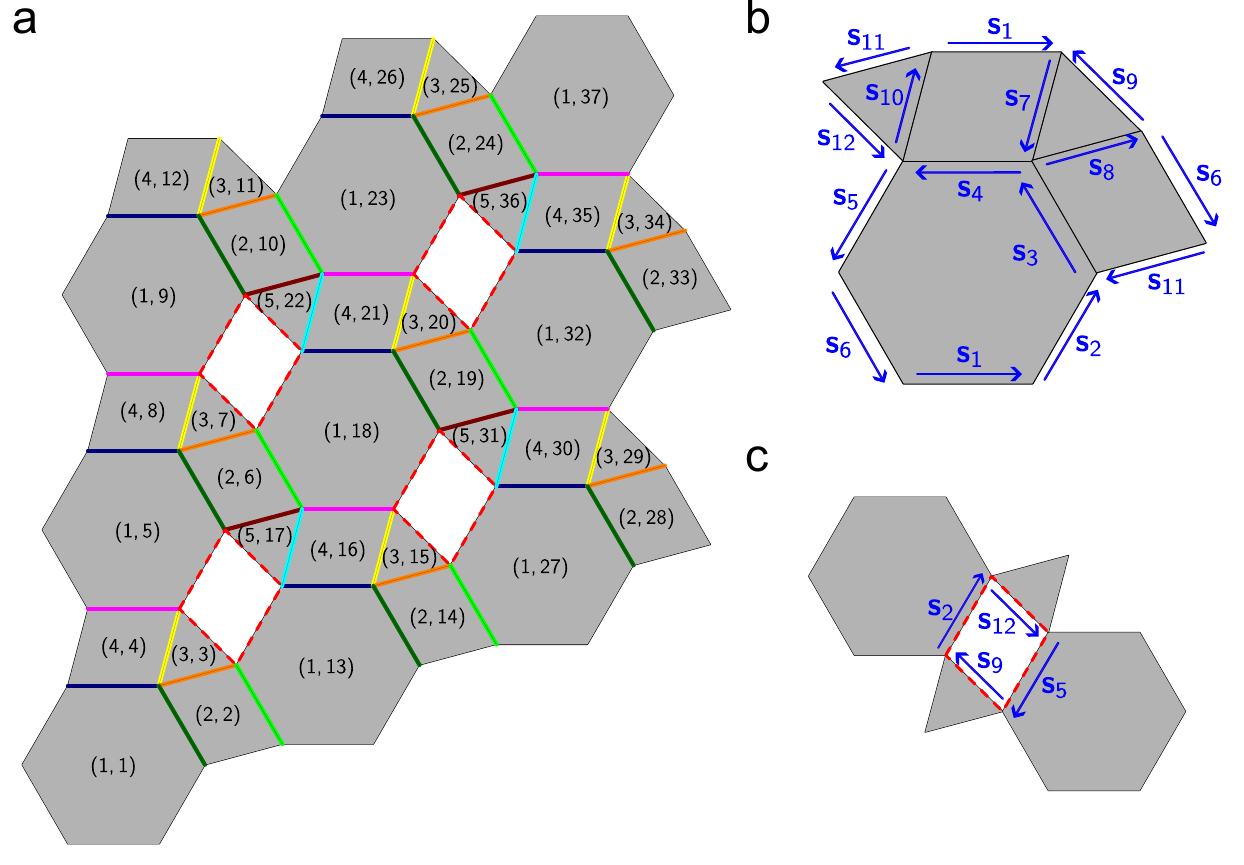}
    \caption{\textit{Notation for the kinematic analysis of a non-planar krigami pattern: (a) the panel labels; (b) the edge vectors that repeat through the pattern; (c) the edge vectors that make up the four slits. } }
    \label{fig:AnalyzeNonPlanarKiri}
\end{figure}

\vspace*{0.2cm}

\noindent \textbf{Uniqueness of the bulk mechanism.} We now show that the pattern in Fig.\;4b of the main text has exactly one mechanism, the single DOF bulk mechanism. Fig.\;\ref{fig:AnalyzeNonPlanarKiri}a shows this pattern --- it is formed by the unit cell in Fig\;\ref{fig:DesignNonPlanarKiri}c, with the boundary pruned to eliminate any obvious boundary floppy modes. As we have already established that this pattern has a bulk mechanism, our goal  is to demonstrate its uniqueness: We do so by a linear   analysis of all of the compatibility conditions of this pattern. 

To set up this calculation, apply  rotation $\mathbf{R}_{i,j} \in SO(3)$ to the panels  labeled by $(i,j)$, as  indicated in Fig.\;\ref{fig:AnalyzeNonPlanarKiri}a. Using this figure combined with the labeling of the edge vectors in Fig.\;\ref{fig:AnalyzeNonPlanarKiri}b, these panel  rotations need to satisfy the crease compatibility conditions
\begin{equation}
\begin{aligned}\label{eq:compatCreases}
   &(\text{6 magenta creases:})  && \mathbf{R}_{i,j} \mathbf{s}_1 = \mathbf{R}_{i',j'} \mathbf{s}_1, \\
   &(\text{8 dark green creases:})  && \mathbf{R}_{i,j} \mathbf{s}_3 = \mathbf{R}_{i',j'} \mathbf{s}_3, \\
   &(\text{8 dark blue creases:})  && \mathbf{R}_{i,j} \mathbf{s}_4 = \mathbf{R}_{i',j'} \mathbf{s}_4, \\
   &(\text{6 light green creases:}) &&\mathbf{R}_{i,j} \mathbf{s}_6 = \mathbf{R}_{i',j'} \mathbf{s}_6,  \\
   &(\text{8 yellow creases:})  && \mathbf{R}_{i,j} \mathbf{s}_7 = \mathbf{R}_{i',j'} \mathbf{s}_7, \\
    &(\text{8 orange creases:})  && \mathbf{R}_{i,j} \mathbf{s}_8 = \mathbf{R}_{i',j'} \mathbf{s}_8, \\
    &(\text{4 light blue creases:})  && \mathbf{R}_{i,j} \mathbf{s}_{10} = \mathbf{R}_{i',j'} \mathbf{s}_{10}, \\
    &(\text{4 maroon creases:})  && \mathbf{R}_{i,j} \mathbf{s}_{11} = \mathbf{R}_{i',j'} \mathbf{s}_{11}, 
\end{aligned}
\end{equation}
where  $(i,j)$ and $(i',j')$ indicate the two panels that form each crease in this calculation. Likewise, in comparing Fig.\;\ref{fig:AnalyzeNonPlanarKiri}a with Fig.\;\ref{fig:AnalyzeNonPlanarKiri}c, the rotations need to satisfy the four loop compatibility conditions 
\begin{equation}
\begin{aligned}\label{eq:compatLoops}
    &\mathbf{R}_{1,5} \mathbf{s}_2 + \mathbf{R}_{5,17} \mathbf{s}_{12} + \mathbf{R}_{1,13} \mathbf{s}_5 + \mathbf{R}_{3,3} \mathbf{s}_9 = \mathbf{0}, && \mathbf{R}_{1,9} \mathbf{s}_2 + \mathbf{R}_{5,22} \mathbf{s}_{12} + \mathbf{R}_{1,18} \mathbf{s}_5 + \mathbf{R}_{3,7} \mathbf{s}_9 = \mathbf{0}, \\
    & \mathbf{R}_{1,18} \mathbf{s}_2 + \mathbf{R}_{5,31} \mathbf{s}_{12} + \mathbf{R}_{1,27} \mathbf{s}_5 + \mathbf{R}_{3,15} \mathbf{s}_9 = \mathbf{0}, && \mathbf{R}_{1,23} \mathbf{s}_2 + \mathbf{R}_{5,36} \mathbf{s}_{12} + \mathbf{R}_{1,32} \mathbf{s}_5 + \mathbf{R}_{3,20} \mathbf{s}_9 = \mathbf{0}. 
\end{aligned}
\end{equation}
These  56 equations express all compatibility conditions parameterizing a mechanism deformation of the pattern.

We confirm that the bulk mechanism is unique as follows: Notice that the $i = 1,\ldots, 5$ index in the $(i,j)$ labeling scheme in Fig.\;\ref{fig:AnalyzeNonPlanarKiri}a repeats for like panels, e.g., the hexegons all correspond to $i = 1$.  The bulk mechanism is given by rotations $\mathbf{R}_{i,j} = \mathbf{R}^{\pm}_i(\omega)$ for all $i = 1,\ldots, 5$  and all $j$, where $\omega$ parametrizes the mechanism motion and $\pm$ indicates its mountain-valley assignment (recall the discussion below (\ref{eq:R2Param})). This prescription solves all the compatibility conditions in (\ref{eq:compatCreases}) and (\ref{eq:compatLoops}). We perturb it via general infinitesimal rotations  by writing
\begin{align}
    \mathbf{R}_{i,j} \approx \mathbf{R}_{i}^{\pm}(\omega) \Big( \mathbf{I} + (\mathbf{w}_j \times) \Big) \quad \text{ for all $i$ and $j$}  
\end{align}
for some $\mathbf{w}_1,\ldots, \mathbf{w}_{37} \in \mathbb{R}^3$.
Substituting this perturbation into (\ref{eq:compatCreases}) and (\ref{eq:compatLoops}) leads to a linear set of equations at leading order of the form 
\begin{align}
    \mathbf{C}^{\pm}(\omega) \mathbf{w} = \mathbf{0}
\end{align}
where  $\mathbf{w} \in \mathbb{R}^{111}$ stacks the $\mathbf{w}_j$ vectors above into any array and  $\mathbf{C}^{\pm}(\omega) \in \mathbb{R}^{168\times 111}$ is a compatibility matrix that depends  on the design vectors $\mathbf{s}_1, \ldots, \mathbf{s}_{12}$ and the bulk mechanism kinematics through $\mathbf{R}_1^{\pm}(\omega), \ldots, \mathbf{R}_5^{\pm}(\omega)$. 

We have numerically evaluated the  null-space of $\mathbf{C}^{\pm}(\omega)$ for a representative sweep of discrete values of $\omega$ and for each mountain-value assignments (We do not include the  flat state $\omega = 0$ in this calculation  because it  has many infinitesimal mechanisms that have no non-linear analogue.) In each case, we have found that the null-space of $\mathbf{C}^{\pm}(\omega)$ is of dimension $4$. Three of the dimensions correspond to an overall rigid rotation of the entire pattern, while the fourth is the infinitesimal perturbation of the bulk mechanism. This analysis justifies the assertion from the main text that the pattern's mechanism has a single DOF.

\section{3D metamaterials }

\begin{figure}[h!]
    \centering
    \includegraphics[width=0.8\linewidth]{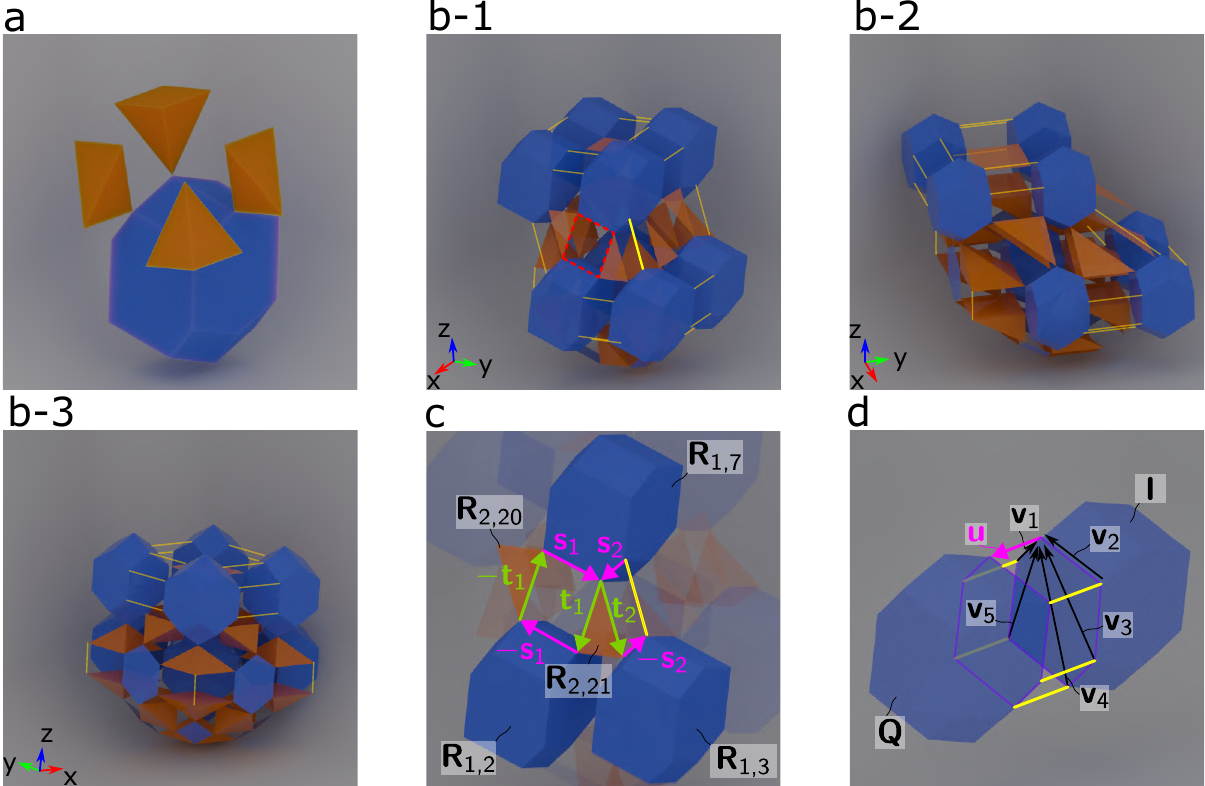}
    \caption{\textit{Infinitesimal mechanism analysis for a 3D metamaterial. 
    (a) Unit cell of the designed pattern, consisting of a blue dodecahedron from the original space tiling and orange polyhedra denoting the added blocks.
    (b) ``Extended unit cell": Unit cell together with its surrounding connected blocks used in the infinitesimal analysis.
    The red parallelogram highlights an example of the slit condition, and the yellow lines represent the length-preservation constraints. (b-1), (b-2), and (b-3) show three different viewing perspectives.
    (c) Enlarged view of the highlighted slit and edge constraints in (b-1), illustrating the local loop compatibility condition employed in the kinematic analysis. 
    (d) Two neighboring dodecahedra connected by six bars are shown to emphasize the local compatibility constraints; $\mathbf{v}_i$ ($i=1,\dots,5$) denote the five vectors used in the rigidity proof of the dodecahedron, and $\mathbf{u}$ denotes the edge vector of the bar connected to the fixed adjacent block/..}}
    \label{fig:3DMetamaterialAnalysis}
\end{figure}

As a final demonstration  of our design principles, we highlight a 3D mechanism-based metamaterial. Fig.\;\ref{fig:3DMetamaterialAnalysis}a shows the  unit cell of  the metamaterial in Fig.\;4c of the main text. It  is obtained from a space tiling of a dodecahedron composed of 4 hexagonal faces and 8 rhombic faces. After separating the  tiling using a tensor $\mathbf{D} \in \mathbb{R}^{3\times3}$ (such that $\min_{\mathbf{e} \in \mathbb{S}^2}| \mathbf{D} \mathbf{e}| > 1$ and $\det \mathbf{D} > 1$), we trace out the convex hull of
the vertices that were initially connected in the space tiling to obtain a new set of polyhedra, the orange blocks in the figure. A key  feature of this pattern, and  any  pattern generated by separating a space tiling in this fashion, is that every loop formed by connecting adjacent panels is a parallelogram. This property is inherent in the construction of the pattern, just as in the 2D recipe. As we show below, it is what allows them to have a mechanism-based response.

The structure on display in Fig.\;\ref{fig:3DMetamaterialAnalysis}b is an   ``extended unit cell", constructed with the help of the visualization software FreeCAD using the following procedure: 1) first add all the building blocks surrounding the cell that connect directly to the cell; 2) then add the building block faces that  constrain the added blocks  kinematically; 3) finally, identify all edges in this system whose lengths need to be preserved under a mechanism deformation. The purpose of this extension is  to  emphasize/catalog the compatibility conditions    associated to a single cell when the overall pattern undergoes a mechanism  deformation. All told, the extended unit cell comprises 8 dodecahedron blocks and 6 additional polygonal faces of such blocks (in blue), 18 polyhedra blocks and 5 additional polygonal faces of such blocks (in orange), and 22 length preserving edge constraints (in yellow). Fig.\;\ref{fig:3DMetamaterialAnalysis} b-1, b-2 and b-3 show the interacting blocks from different viewing perspectives.

We now apply rotations to each block to study the mechanisms  of this cell. The rotations associated with the blocks and faces originating from the dodecahedral space tiling are labeled as $\mathbf{R}_{1,j} \in SO(3)$, $j = 1,\ldots, 14$; the corresponding rotations of the orange blocks and faces introduced by the added polyhedra are denoted by $\mathbf{R}_{2,j}$, $j  = 15, \ldots, 37$.   Fig.\;\ref{fig:3DMetamaterialAnalysis}c illustrates one of the loop compatibility conditions on the rotations, i.e., 
\begin{align}\label{eq:slitCompat3D}
(\mathbf{R}_{1,2} - \mathbf{R}_{1,7}) \mathbf{s}_1  + (\mathbf{R}_{2,20} - \mathbf{R}_{2,21}) \mathbf{t}_1 = \mathbf{0}.
\end{align} Fig.\;\ref{fig:3DMetamaterialAnalysis}c also illustrates one of the length preserving ``edge" constraints  present in this extended unit cell, i.e., 
\begin{align}\label{eq:edgeCompat}
|\mathbf{t}_2|^2 = |\mathbf{R}_{1,7}\mathbf{s}_2+\mathbf{R}_{2,21}\mathbf{t}_2-\mathbf{R}_{1,3}\mathbf{s}_2|^2.
\end{align}
Notice that $\mathbf{R}_{1,2} = \mathbf{R}_{1,3} =  \mathbf{R}_{1,7} = \mathbf{I}$ and $\mathbf{R}_{2,20} = \mathbf{R}_{2,21} = \mathbf{R}$ solves (\ref{eq:slitCompat3D})  and  (\ref{eq:edgeCompat}) for any rotation $\mathbf{R} \in SO(3)$. In fact,  every slit in this extended cell is a parallelogram connecting adjacent blue and orange blocks and every edge constraint corresponds to such a parallelogram slit in the overall pattern. Thus,  prescribing the  rotations as  
\begin{align}
\mathbf{R}_{1,j} = \mathbf{I}, \quad j = 1,\ldots, 14  \quad \text{ and } \quad \mathbf{R}_{2,j}  = \mathbf{R} \quad j = 15,\ldots, 37
\end{align}
solves all the compatibility conditions of the extended cell. As this prescription is also periodic, it can be tessellated to produce a mechanism deformation of the overall pattern, parameterized by the 3 DOFs of the rotation  $\mathbf{R}$. 

Naturally,  we would like to know whether there are any additional mechanism deformation of this pattern. Thus, following the same procedure as in Section \ref{sec:infiniteAnalysis}, we study linear perturbations of this periodic mechanism via 
\begin{align}\label{eq:perturb3D}
\mathbf{R}_{1,j} \approx \mathbf{I} + (\mathbf{w}_j \times),   \quad  j =1,\ldots, 14, \quad \mathbf{R}_{2,j} \approx \mathbf{R} \big( \mathbf{I} + (\mathbf{w}_j \times) \big) , \quad  j = 15, \ldots, 37. 
\end{align}
Again, with the help of  FreeCAD, we have carefully accounted for all the loop compatibility conditions and all edge constraints in the extend cell --- there are 100 parallelogram slit loop conditions and 22 line length preservation conditions in total. Enumerating all of these conditions under the ansatz (\ref{eq:perturb3D}) leads to a linear compatibility equation of the form 
\begin{align}
\mathbf{C}(\mathbf{R}) \mathbf{w} = \mathbf{0}
\end{align} 
where, as before, $\mathbf{w} \in \mathbb{R}^{111}$ stacks the vectors $\mathbf{w}_{j}$ into an array and $\mathbf{C}(\mathbf{R}) \in \mathbb{R}^{322 \times 111}$ is the corresponding linear compatibility matrix, which depends on $\mathbf{R}$. Again, we have evaluated the null-space of $\mathbf{C}(\mathbf{R})$ for a representative sweep of numerical values of $\mathbf{R}$. This null-space is seven dimensional for each evaluation of $\mathbf{R}$ --- three DOFs for an overall rotation of the pattern and three corresponding to an infinitesimal perturbation of the bulk mechanism and one extra twist-like infinitesimal mechanism.

The unexpected additional  infinitesimal mechanism is parameterized as follows: after fixing two adjacent blocks ($\mathbf{w}_1=  \mathbf{w}_{15} =  \mathbf{0}$) to eliminate the overall rotation and bulk mechanism, the remaining vector in the null-space of $\mathbf{C}(\mathbf{R})$  describes a twist-like infinitesimal mode of each blue dodecahedron relative to the fixed central dodecahedron of the unit cell. We show that this infinitesimal mechanism does not extend to a fully non-linear mechanism by isolating  the subset of the unit cell in  Fig.\;\ref{fig:3DMetamaterialAnalysis}d  (two dodecahedrons, connected by six length preserving edges) and proving that it cannot maintain this kinematics at large deformations without violating some of the compatibility conditions for a mechanism deformation.

In this setup, we are free to fix one dodecahedron  and one edge (labeled $\mathbf{u}$ in the figure), so as to be consistent with the twist-like infinitesimal mode. The other dodecahedron is then allowed to rotate via some $\mathbf{Q} \in SO(3)$. If a nonlinear mechanism exist, then $\mathbf{Q}$  must satisfy  
\begin{equation}
    \begin{aligned}
        |\mathbf{u}|^2 &= \left| \mathbf{v}_i + \mathbf{u} - \mathbf{Q}\mathbf{v}_i \right|^2, \qquad i = 1,\ldots,5.
    \end{aligned}
\end{equation}
 to preserve the lengths of the other five edges in the figure. 
These expressions simplify to 
\begin{equation}
\label{eq:NonCompaCons}
(\mathbf{Q} - \mathbf{I})\mathbf{v}_i \cdot (\mathbf{v}_i + \mathbf{u}) = 0,\qquad i = 1,\ldots,5.
\end{equation}
We show that  the infinitesimal mode is obstructed at second order. Indeed, using the matrix exponential, a generic rotation is parameterized to second order by 
\begin{align}
\mathbf{Q} = \exp \big( \boldsymbol{\omega} \times \big)
= \mathbf{I} + \boldsymbol{\omega}\times + \frac{1}{2}(\boldsymbol{\omega}\times)^{2} + \mathcal{O}(|\boldsymbol{\omega}|^3).
\end{align}
The nonlinear residual functions of our compatibility constraints are thus
\begin{equation}
r_i(\boldsymbol{\omega}) 
:= 
\big(\exp \big(\boldsymbol{\omega} \times \big) - \mathbf{I}\big)\mathbf{v}_i 
\cdot 
(\mathbf{v}_i + \mathbf{u}),
\qquad i = 1,\ldots,5.
\end{equation}
Substituting $\boldsymbol{\omega}=\varepsilon\boldsymbol{\omega}_1+\tfrac{1}{2}\varepsilon^{2}\boldsymbol{\omega}_2$ and collecting powers of $\varepsilon$ gives 
\begin{align}
r_i\big(\varepsilon\boldsymbol{\omega}_1+\tfrac{1}{2}\varepsilon^{2}\boldsymbol{\omega}_2\big)
&=
\varepsilon  \big((\boldsymbol{\omega}_1\times\mathbf{v}_i)\cdot(\mathbf{v}_i+\mathbf{u})\big)
+\varepsilon^{2}\frac{1}{2}\Big( \boldsymbol{\omega}_2\times\mathbf{v}_i + \boldsymbol{\omega}_1\times(\boldsymbol{\omega}_1\times\mathbf{v}_i) \Big)
\cdot(\mathbf{v}_i+\mathbf{u})
+\mathcal{O}(\varepsilon^{3})
\end{align}
for $i = 1,\ldots,5$. Observe that $\boldsymbol{\omega}_1 = \lambda \mathbf{u}$ for $\lambda \in \mathbb{R}$ parameterizes the infinitesimal twist-like mode, as it nullifies the residual at $O(\varepsilon)$. Substituting this parameterization back into the residual gives the equations at $O(\varepsilon^2)$: 
\begin{align}\label{eq:secondOrder}
 (\boldsymbol{\omega}_2 \times \mathbf{v}_i)  \cdot (\mathbf{v}_i + \mathbf{u}) = - \lambda^2  \mathbf{u} \times ( \mathbf{u} \times \mathbf{v}_i)  \cdot ( \mathbf{v}_i + \mathbf{u}) ,\quad i=1,\ldots, 5.
\end{align} 
Let $\boldsymbol{\omega}_2 = \lambda^2 \hat{\boldsymbol{\omega}}$. After applying various cross-product  identities, the expression in (\ref{eq:secondOrder}) becomes  
\begin{align}\label{eq:linear-cancel}
(\mathbf{v}_i \times \mathbf{u}) \cdot \hat{\boldsymbol{\omega}}  = |\mathbf{u} \times \mathbf{v}_i|^2 , \quad i =1,\ldots,5.
\end{align}
Equation~\eqref{eq:linear-cancel} constitutes a $5 \times 3$ linear system for $\hat{\boldsymbol{\omega}}$. We have evaluated numerically that this system admits no solution for our design.  Consequently, the residual remains of order $\mathcal{O}(\varepsilon^2)$ and is strictly nonzero for all sufficiently small nonzero $\varepsilon$. This implies that the infinitesimal mode is obstructed at second order, confirming the assertion in the main text that this 3D metamaterial has a 3 DOF bulk mechanism and no other nonlinear floppy modes.

\bibliography{PRLRef1}